\documentclass[article]{revtex4}

\usepackage[latin1]{inputenc}
\usepackage{graphicx}
\usepackage{dcolumn}
\usepackage{bm}
\usepackage{color}

\newcommand{\ket}[1]{|#1\rangle}
\newcommand{\bra}[1]{\langle#1|}

\newcommand{\g}{|g\rangle}
\newcommand{\s}{|s\rangle}
\newcommand{\e}{|e\rangle}
\newcommand{\gb}{|\mathbf{g}\rangle}
\newcommand{\sbN}{|\mathbf{s^{N_a}}\rangle}
\newcommand{\sbold}{|\mathbf{s^1}\rangle}

\newcommand{\0}{|0\rangle}
\newcommand{\1}{|1\rangle}
\newcommand{\+}{|+\rangle}
\newcommand{\m}{|-\rangle}
\newcommand{\Cbas}{\{\0,\1 \}}
\newcommand{\Hbas}{\{\+,\m \}}

\begin{document}

\title{Quantum interrogation logic gates.}
\author{Juan Carlos Garc\'ia-Escart\'in}
\author{Pedro Chamorro-Posada}
\affiliation{Departamento de Teor\'ia de la Se\~{n}al y Comunicaciones e Ingenier\'ia Telem\'atica. Universidad de Valladolid. Spain.}
\date{\today}
\email[Contact electronic address:]{juagar@tel.uva.es}
\begin{abstract}
Quantum interrogation can be used as a basic resource for quantum information. This paper presents its applications to entanglement creation and to optical quantum computation. The starting point will be a photon to particle quantum interrogation CZ gate, from which three families of optical CNOT gates will be derived. These gates are nondestructive and can work with probabilities arbitrarily close to 1. A possible experimental implementation with atomic ensembles is also discussed.
\end{abstract}  
\maketitle

\section{Introduction}
Quantum computation and quantum communication promise new applications that can go beyond those of classical computing \cite{NC00}. In order to fully exploit the power of quantum information processing, scalable quantum computers need to be built. From all the candidates, optical quantum computation offers a qubit which is particularly well suited for long distance communication, has long coherence time and can be manipulated with many existing instruments. Interaction of different optical qubits, however, has proven to be extremely elusive. Quantum computing with photons has received a great deal of interest after the notable demonstration that quantum computation can be done with linear optics elements alone \cite{KLM01}. This has led to many proof of principle gates \cite{OPW03}, which are probabilistic by nature and, in most cases, destructive. In this paper, we show an alternative optical CNOT gate based on quantum interrogation of quantum particles. 

The interferometric setup of quantum interrogation can be easily converted into a CZ gate between photons and particles, which can be part of a gate teleportation scheme that gives an optical CNOT gate. The particle acts as an ancillary system and can be discarded afterwards. These results are valid for the interrogation of any quantum object that can absorb, scatter, or in any other way trigger the detection of a single photon located in its vicinity with a high enough probability. 

We will first show how quantum interrogation can produce different entangled states of light. This entanglement can later be used to create quantum gates with gate teleportation \cite{GC99}. We present three families of optical CNOT gates based on quantum interrogation. These gates are nondestructive and work with a probability arbitrarily close to 1 in an ideal operation with no losses. The schemes are valid for any matter-to-light CZ gates and could work even for matter qubits with short coherence times. 

Section \ref{gates} enumerates the different quantum gates that appear in the discussion and establishes our notation. Section \ref{LOQC} sets down the basic concepts of linear optical quantum computation. Section \ref{QI} refreshes the concepts of quantum interrogation that are behind the proposals in this paper. In section \ref{IFentang}, we introduce interaction-free entanglement creation schemes, which are the basic building blocks of the following applications. Generation schemes for optical Bell and W states are discussed. Section \ref{QMastel} shows a quantum teleportation circuit that can be built from the proposed CZ gates and which can be interpreted as a quantum memory. Section \ref{3vers} uses all the previous work to propose alternative versions of optical CNOT gates based on gate teleportation and QI entangling gates. Section \ref{implement} covers the physical implementation aspects, including a proposal for QI gates with atomic ensembles. Section \ref{discussion} reviews the most important features of the new scheme and discusses the results. 

\section{Gates and notation}
\label{gates}
The operations of quantum computing can be described as quantum circuits, which are formed by different quantum gates. An $n$-qubit quantum gate can be defined as a system that performs a determined operation on $n$ input qubits so that for each input value there is a defined associated output. Superpositions of different input states will produce the corresponding superpositions of output states. 

We will need three different single qubit gates: NOT, Hadamard and Z gates. The NOT, or X, gate generalizes the classical NOT gate and flips the value of the qubit it is acting on. Usually, this is written as $X\ket{x}=\ket{x\oplus 1}$, where $\oplus$ is used to account for a XOR, or modulo 2 addition, operation. The operation of the H gate can be seen from its effect on the states of the computational basis, $\Cbas$, with $H\ket{0}=\ket{+}=\frac{\ket{0}+\ket{1}}{\sqrt{2}}$ and  $H\ket{1}=\ket{-}=\frac{\ket{0}-\ket{1}}{\sqrt{2}}$. Superpositions of $\0$ and $\1$ result in the corresponding superpositions of $\+$ and $\m$. We will stress the meaning of the H gate as a base conversion operation from the $\Cbas$ into the $\Hbas$ basis. The Z gate performs a sign shift when the value of the qubit is $\1$ and does nothing otherwise. The operation can be written as $Z\ket{x}=(-\!1)^x\ket{x}$. The most widely used two qubit gates are the CNOT gate, of operation $CNOT\ket{x}\ket{y}=\ket{x}\ket{x\oplus y}$ and the CZ gate, for which $CZ\ket{x}\ket{y}=(-1)^{xy}\ket{x}\ket{y}$. All these gates are their own inverse. 

\begin{figure}[ht!]
\centering
\includegraphics[scale=0.8]{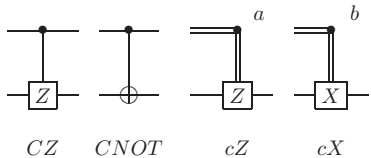}
\caption{Circuit representation of the used controlled gates.\label{Cgates}} 
\end{figure}

We will use both quantum and classically controlled quantum gates. We follow the usual notation that represents a control by a dot if the gate is activated by a $\ket{1}$ state or a blank circle if it is the $\ket{0}$ state that activates the gate operation. A gate with more than one control is only applied if all the conditions are simultaneously met, and will act as the identity operator otherwise. Classically controlled U gates will be represented as cU. The small c indicates the control is classical instead of quantum and the pictorial representation includes the usual double lines for classical bits. Figure \ref{Cgates} contains the representation of these controlled gates in the circuit model.

A set of gates which, when correctly combined, can be used to implement any logic function is called a universal set. The set of all single qubit gates and the CNOT gate is one of those universal sets \cite{BBC95}.

\section{Linear optics quantum computation}
\label{LOQC}
Linear optics quantum computation, LOQC, is probably the simplest quantum computing implementation proposal. In LOCQ, the qubits are photons and the quantum gates are standard optical elements that are well characterized and already in wide use. 

In our discussion, the optical qubits will be encoded in dual-rail representation. In dual-rail, $\0$ and $\1$ are encoded by a single photon in a different orthogonal optical mode. The modes can be orthogonal polarizations or different paths. The most usual encodings are horizontal and vertical polarization, so that $\0= \ket{H}$ and $\1=\ket{V}$, and spatial encoding, where $\ket{0}=\ket{10}$ and $\1=\ket{01}$ and the photon goes through different paths for each value.

Spatially encoded qubits will be our default representation, but there will be occasional comments on the equivalent polarization encoding. Any other system showing the superposition of different modes could be used instead. 

For this dual-rail encoding, any single qubit transformation can be induced by using only beamsplitters and phase shifters \cite{KLM01}. Nevertheless, it is exceedingly difficult to make two photons interact. Current optical CNOT gate proposals can achieve the required interaction with varying degrees of success \cite{KMN07}, but there are still no simple optical CNOT gate models.

We will assume that perfect arbitrary single qubit gates are available for our optical qubits and concentrate on the realization of an efficient CNOT gate that would complete a universal set of gates for scalable quantum computation. 

Another important element for the construction of a quantum computer is a mechanism for the storage and reading of the qubits: a quantum memory. A quantum memory must allow an easy conversion between flying and stationary qubits. Different potential realizations for quantum memories have been proposed \cite{FL02,JSC04}, but most of them require difficult to keep conditions or high precision control. We will also give a model for quantum interrogation memories which can relax some of the demands.

It is the lack of efficient CNOT gates and quantum memories that has hindered the development of linear optical quantum computers. With our proposal, we extend the capabilities of linear optics schemes and offer an alternative physical realization for a quantum computer.

\section{Quantum interrogation}
\label{QI}
Quantum interrogation, QI, offers a perspective on measurement and interaction that is profoundly different from the classical picture. QI is a form of interaction-free measurement, or IFM. IFM allows to obtain information on the state of an object without interacting with it in any meaningful classical way \cite{Dic81,Vai01,Tse98}. Quantum Interrogation methods combine concepts from the Zeno effect and IFM to increase the efficiency of the schemes.

\begin{figure}[ht!]
\centering
\includegraphics[scale=0.7]{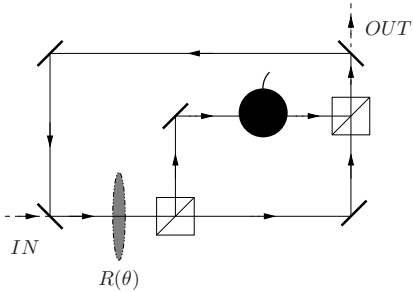}
\caption{Quantum interrogation in an interferometric-like setup with two polarizing beamsplitters and a polarization rotator.\label{interferometer}} 
\end{figure}

We will follow the ``polarization interferometer'' setup of the experiment of Kwiat \cite{KWM99}, which can be explained from Figure \ref{interferometer}.

In the setup, two polarizing beamsplitters, PBS, separate the horizontally and vertically polarized components of an input photon. Each component is directed to a different arm. The horizontal component, state $\ket{H}$, is directed into the lower arm, while vertically polarized light, in $\ket{V}$, goes to the upper arm. The oval represents a polarization rotator. 

Although in the setup there is no interference proper at the output of the second PBS, we will call the system formed by the PBSs and the polarization rotator an interferometer. The arms of the setup will be identified with the arms of an interferometer. Here, however, interference happens at the polarization rotator instead of at the place where the arms are reunited. 

In the quantum interrogation scheme, the upper arm of the interferometer can contain a bomb so sensitive that it can detect a single photon. If a photon is present in the upper arm of the interferometer, it is detected and the bomb explodes in response. 

In order to test for the presence the bomb, an input photon with horizontal polarization, in state $\ket{H}$, is introduced into the interferometer, where it is kept until it has completed $N$ cycles inside the loop of Figure \ref{interferometer}.

For an empty interferometer with no bomb, the polarization is rotated by an angle $\theta$ each cycle. At the end of the procedure there is a global rotation of $N\theta$. For $\theta=\frac{\pi}{2N}$, a horizontally polarized probe photon in $\ket{H}$ ends in a vertically polarized state $\ket{V}$.  

The presence of a bomb, on the other hand, would inhibit the evolution at each step. We will consider the bomb to act as a projective measurement. The photon's state is rotated from $\ket{H}$ into $\cos{\theta}\ket{H}+\sin{\theta}\ket{V}$. For a small enough angle the photon is projected back to $\ket{H}$ with high probability.

The continuous measurement either results in an $\ket{H}$ state, or triggers an explosion. The preservation of the input state happens with a probability $\cos^2\theta \approx (1-\frac{\theta^2}{2})^2$ in each step, with a total probability of $\cos^{2N}\theta \approx 1-N\theta^2$. If we choose a $\theta$ proportional to $\frac{1}{N}$, this probability can be arbitrarily close to 1, when N is large enough. 

For $\theta=\frac{\pi}{2N}$, the output tells us whether there is a bomb, for output $\ket{H}$, or not, for output $\ket{V}$, yet no photon can have touched the bomb, or it would have exploded.

We will consider the case where $\theta=\frac{\pi}{N}$. Here, in absence of the bomb, $\ket{\psi_{out}}=-\ket{H}$. The output photon is still horizontally polarized, but a sign shift has taken place. For an isolated photon, the addition of a global phase to the state has no measurable effect. However, if the photon in the interferometer is part of a superposition with a quantum bomb, the phase shift with respect to the other terms can have important effects. This relative phase shift will be the key in the creation of entanglement and quantum gates based on QI.

\section{Quantum interrogation gates for the creation of entanglement}
\label{IFentang}
For a superposition of bomb states, the conditional effect on the state of a photon can be translated into entanglement between the bomb and the photon. The QI setup is then used as an entangling photon-particle gate. This creation of entanglement will be the origin of all the capabilities of our quantum interrogation applications.

\subsection{CZ gate between light and particles}
\label{QICZ}
We will work with two-state systems. The photonic qubit will be encoded in dual-rail. The bomb qubit will be encoded in the position of a particle, which can be either blocking the path through the interferometer or elsewhere. We will label as $\0$ the state of a bomb blocking the interferometer (closed interferometer) and as $\1$ the bomb state for which the interferometer is free (open). This notation makes no assumption on the interrogated object and is more general than a notation based on the absence, $\0$, or presence of the particle, $\1$. Additionally, it allows a more intuitive interpretation in terms of well-known standard qubit gates.

\begin{figure}[ht!]
\centering
\includegraphics[scale=0.75]{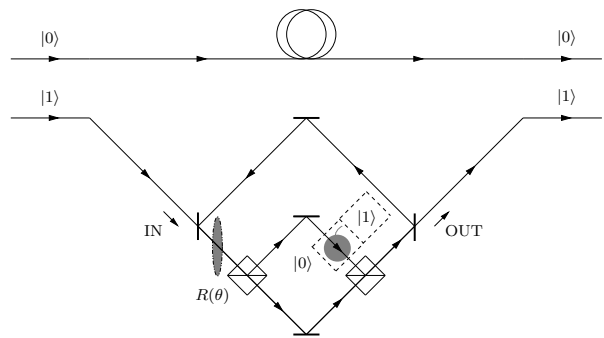}
\caption{CZ gate for a quantum bomb (shaded) and a dual-rail qubit.\label{CZ}} 
\end{figure}

If a photon goes through a QI cycle through our interferometer, there will be a $\pi$ phase shift if the particle is in $\1$ and no effect for the $\0$ state. For a photonic dual-rail qubit where the photon mode that corresponds to $\1$ is introduced in the QI interferometer, the relative phase shift between components will appear only when the interferometer is free of particles (state $\ket{1}$). The operation is exactly that of a CZ gate; the input state is conserved except for the $\ket{11}$ state, in which case a $\pi$ phase shift with respect to the other states occurs. The configuration shown in Figure \ref{CZ} produces this CZ gate effect. The upper path is the input of the logical $\ket{0}$ mode and the lower line corresponds to the photon mode of the logical $\ket{1}$. The bomb has been shaded as a reminder that it can be in a superposed state of being and not being there.

The sign change will only appear in the states of the superposition that have both a photon in the lower port and an empty interferometer. 

Although it is not strictly necessary, the interpretation becomes clearer if we assume a continuous observation of the particle so that the state for which an absorption takes place (bomb explosion) is amplified to the classical level. The continuous measurement process distinguishes only between photon absorption, or scattering, and the rest of the cases.

We can see the process as a continuous projection into the $\ket{A}\bra{A}$ and $I-\ket{A}\bra{A}$ subspaces, with $\ket{A}$ the state resulting from the absorption, or scattering, of the photon. If $N \rightarrow \infty$, the effect of rescaling the probability amplitudes is negligible and the measurement will not affect the part of the superposition with photon state $\0$.

Once there is a CZ gate available, a CNOT gate can be obtained just by adding two H gates on the target qubit before and after the CZ gate. For optical qubits this amounts to two extra $50\%$ beamsplitters. It is equally possible to have a control optical qubit and a target particle qubit. In our proposal, we will keep the operations on the particle as simple as possible. The central part of the proposal will be optical. The part of the bomb could be played by any object able to absorb a photon and to show superposition. The only additional requisites in our scheme are state creation and measurement in the $\Hbas$ basis.

\subsubsection{Comparison with other quantum interrogation proposals}
\label{comp}
The presented CZ gate can also help to overcome the weaknesses of previous proposals for entanglement creation with QI of quantum particles. First, it can be applied on arbitrary input qubits for both the photon and the particle. Some of the proposed QI gates are based on partial CNOT gates \cite{GWM02} or have input ports that cannot be used \cite{Azu03,Azu04}. Second, it is highly symmetric, so control and target can be easily interchanged. Third, the operation in the ideal regime can be made arbitrarily close to 1. Later, this will allow us to build optical CNOT gates which are, in principle, deterministic for any input state, in contrast to the schemes of \cite{GWM02,Azu03}, or the nondestructive CNOT gate of \cite{Pav07}, where the erasure of correlations introduces a probabilistic element. This QICZ gate and the optical CNOT gates derived from it are deterministic in the ideal case and probability of failure is, as in most quantum gate proposals, loss or decoherence dependent.

The choice of dual-rail qubits allows for these deterministic setups. Dual-rail encoding was already employed in the deterministic gates of \cite{Azu04}, which still have a restricted input and can only give CNOT gates based on gate teleportation. This restriction forbids some of the simplifications of our direct CNOT proposal. 

This QICZ is a simplification of our previous light-matter CNOT gate, the quantum shutter \cite{GC06a}, in which a particle controls the passage of light through a slit. When the particle is present, the slit is closed, $\0$, and photons are reflected. When there is no particle, we have an open slit, state $\ket{1}$, and the light is not affected. The new description of the QICZ gate gives a simpler implementation of a quantum shutter with only one QI round and two beamsplitters, instead of the previous scheme with two coupled interferometers. Recent proposals of deterministic Zeno gates are also equivalent to our QICZ gate \cite{HM08,WYN08}. Here, however, the stress will be on a light-matter hybrid gate in which the demands on the particle system are as non-restrictive as possible. This way simpler matter systems could be employed. This separates our proposal from the IFM-based deterministic CNOT gate of M\'ethot and Wicker \cite{MW01}. In their gate, the qubits are implemented in atoms inside cavities. These qubits are then coupled with the help of light pulses. In our proposal, we will not need to produce arbitrary superpositions of atomic states and cavity coupling will be avoided if possible. 

The advantages of deterministic operation and the ability to work with arbitrary inputs are also present in the Zeno gates proposed for implementation on coupled optical fibres or microcavities \cite{FJP04}. In those gates, the optical qubits can travel through a two-photon-absorption, TPA, medium. The possibility of the absorption of photon pairs inhibits the evolution for certain input states, while allows it for others. The continuous journey through the medium replaces the discrete number of cycles inside the QI setup. The main drawback of this approach is finding a strong enough non-linear medium so that the TPA can produce a coherent evolution before single photon loss is too high. We will provide the non-linear interaction by means of the particle bomb. The additional complication of the particle system can be compensated by additional benefits which will be shown in the following sections, such as memory schemes, CNOT gates that regenerate the photons, or mechanisms to check for photon loss. 

\subsection{Generation of entangled states}
Modified QI schemes can be used to engineer different kinds of entangled states. In the following sections, we will present different optical CNOT gates based on QI. The combination of CNOT gates with the existing one qubit gates allows to build any possible quantum logic function. Therefore, for a given input state, the production of any entangled state of interest would be straightforward. In this section, however, we are more interested in configurations that can directly produce complex entangled states with simple physical systems. This way, the various entangled states that are a prerequisite for many quantum information applications could be generated with relatively simple equipment. 

All the presented schemes have in common that the entanglement between photons is created by means of ancillary particles. As direct interaction between photons is inaccessible, we recourse to a system that interacts well with light. The whole concept is similar to that of entanglement swapping, where two particles that have never interacted become entangled with the help of entanglement with intermediary systems and measurement \cite{PBW98}.

\subsubsection{Bell states}
\begin{figure}[ht!]
\centering
\includegraphics[scale=0.8]{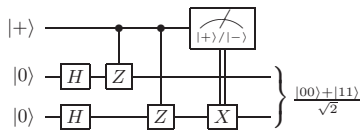}
\caption{Configurable QI gate.\label{entang}} 
\end{figure}

Bell states are the most important example of entangled states and appear in many quantum information applications. With the described CZ gate, it is easy to create the basic Bell state $\frac{\ket{00}+\ket{11}}{\sqrt{2}}$, which allows for a wide range of quantum information protocols. Figure \ref{entang} shows a quantum circuit that can generate a photonic Bell pair from two blank photons. We represent a measurement in the $\{ \ket{+},\ket{-}\}$ basis writing the basis states in the measurement box. 

For this circuit, the two CZ gates can be merged into a single QI stage. We only need to nest two of the interferometers of Figure \ref{interferometer} so that they share the same bomb. We could also forget the postcorrection stage. In that case, the result of the particle measurement would indicate in which Bell state the pair of photons is,  $\frac{\ket{00}+\ket{11}}{\sqrt{2}}$, if $\+$ is found, or  $\frac{\ket{01}+\ket{10}}{\sqrt{2}}$ for $\m$. 

The application of the QICZ gate to Bell state generation is, in itself, an important step towards LOQC. The probabilistic gates of the Knill-LaFlamme-Milburn, KLM, scheme can be more efficiently implemented if, instead of a large number of ancillae and detectors, entangled resource states are employed \cite{PJF01}. Bell pairs can be used to fuel feed-forwardable linear optical CNOT gates, which have already been experimentally demonstrated using parametric down-conversion as the Bell state source \cite{GPW04}. The gates are still probabilistic by nature, but they are non-destructive and it is possible to ascertain whether the gate operation succeeded or failed. The operation can then be repeated as many times as needed until it is correctly performed. QI gates can provide an on-demand source for Bell states.

\subsubsection{Multiple bombs and configurable gates}
\begin{figure}[ht!]
\centering
\includegraphics[scale=0.7]{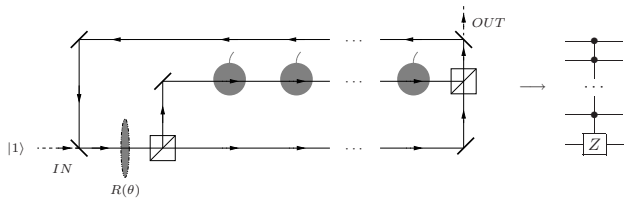}
\caption{QI interferometric setup with multiple bombs.\label{multiple}} 
\end{figure}

If, instead of a single interferometer for many photons, we have an interferometer with more than one particle, we can obtain a CZ gate with multiple control qubits. In the QI scheme of Figure \ref{multiple}, a single bomb is enough to perform a measurement. The only state where all the bombs are out of the path of the photon is $\ket{11\ldots1}$ and only in this case will the sign change occur. If the input is the $\1$ path of a dual-rail qubit, the operation is equivalent to a multiply controlled Z gate. Adding two photonic H gates (two $50\%$ beamsplitters) before and after the interferometer, this gate can be converted into a multiply controlled NOT gate. For instance, for two particles, a Toffoli gate, the generalization of a CNOT gate with two control qubits, can be obtained. 

This configuration can be used to create entanglement between particles and offers interesting possibilities to build configurable gates. For $m$ particles with two position states $\ket{0}$ and $\ket{1}$, arbitrarily defined, the path of the photon might be blocked by the $\ket{0}$ or $\ket{1}$ states. The same particles can, simultaneously, be part of another QI setup. Any of the $2^m$ possible combinations of control, one for every combination of 0s and 1s, can be achieved and configurable devices that depend on the arrangement of the quantum bombs can be built. 

Imagine a ``box'' divided into two halves, each able to hold a particle. The position inside the box can be associated to a logical value and admits a superposition of both $\0$ and $\1$ locations. If there is a particle inside each of a certain number of boxes, the resulting gate can be reconfigured by displacing the boxes which contain the $\0$ and $\1$ states of the bomb.  

\begin{figure}[ht!]
\centering
\includegraphics[scale=0.5]{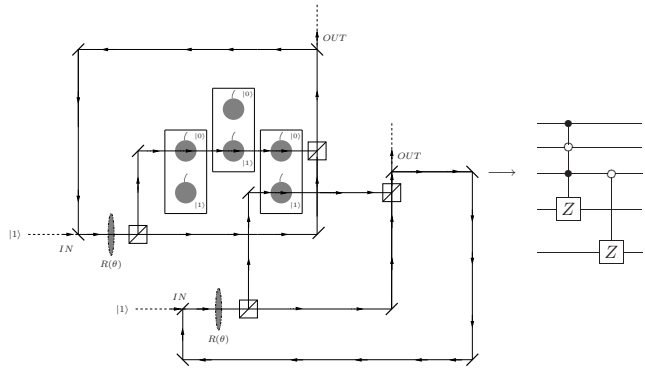}
\caption{Configurable QI gate.\label{confCZ}} 
\end{figure}

Figure \ref{confCZ} shows an example of a more complex QI system and the equivalent quantum circuit that results from connecting the $\ket{1}$ modes of dual-rail qubits to both interferometers. These kinds of gates can help to tailor more elaborate entangled states.  

With relatively simple constructions we can design classically configurable quantum gates and implement different unitary transformations with the same device. One and the same configurable setup can provide any quantum gate from a particular set without recourse to complex gate sequences. This kind of construction will bring closer the realization of a classically controlled quantum compiler that can be programmed to give any desired quantum function. 

Gate standardization is important to be able to reach an economy of scale with more efficient and economical quantum gates and it is the driving reason behind the use of universal sets. Configurable gates constitute a common basic block that can provide a wide range of quantum logic operations and reduce the global gate number. Taking a part of the control into the classical realm makes the programming easier to implement.

\subsubsection{W states}
A W state generator can also be described as a QI setup. W states are another family of entangled quantum states different from the cat or GHZ states, which are the many qubits version of the Bell state $\frac{\ket{00}+\ket{11}}{\sqrt{2}}$. W states cannot be reduced to cat states using only local operations \cite{DVC00}. Among other properties, they show a form of entanglement that is particularly robust against qubit loss. An $M$ qubit W state has the form $\frac{\ket{10\ldots 0}+\ket{01\ldots 0}+\ldots +\ket{00\ldots 1}}{\sqrt{M}}$. From the $M$ qubits of each term, $M-1$ qubits are $\0$ and one is $\1$. The $M$ terms of the superposition correspond to the $M$ possible positions of the $\1$ qubit.

To produce a W state with QI, we can picture a box with $M$ subsections. A particle is then put in a uniform coherent superposition of all the possible positions inside the box, from $\0$ to $\ket{M-1}$. If we have $M$ interferometers with $\theta=\frac{\pi}{2N}$ and each section is blocking a different interferometer, an input state with a horizontally polarized photon for each interferometer will become $\frac{\ket{0}\ket{VHH\ldots H}+\ket{1}\ket{HVH\ldots H}+\ldots+\ket{M-1}\ket{HHH\ldots V}}{\sqrt{M}}$, where only one photon is vertically polarized.

In the same way, in our $\theta=\frac{\pi}{N}$ scheme, $\ket{+}$ input states would transform into $\ket{-}$ only for the corresponding particle state. To obtain an independent optical W state, the entanglement with the particle qudit, here an M-dimensional system, must be erased. The position could be erased by an accurate momentum measurement or by applying the qudit equivalent to the H gate, the Quantum Fourier Transform, followed by a position measurement that will determine the appropriate phase postcorrection. 

\section{Quantum teleportation seen as a memory}
\label{QMastel}
\begin{figure}[ht!]
\centering
\includegraphics[scale=0.9]{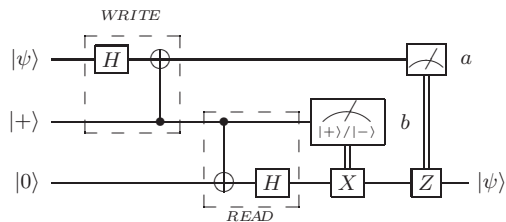}
\caption{Teleportation as a memory write and read cycle.\label{teleportation2}} 
\end{figure}

Quantum teleportation \cite{BBC93} is a powerful primitive in quantum information. The standard quantum teleportation circuit can be rewritten as a memory made from two state transfer circuits. The two steps of teleportation can be separated into a write and a read stage (Figure \ref{teleportation2}) and the intermediate state is considered to be the memory.

\begin{figure}[ht!]
\centering
\includegraphics{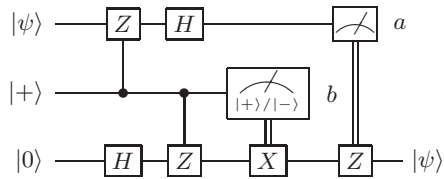}
\caption{Teleportation as a memory with CZ gates.\label{teleportation3}} 
\end{figure}

The circuit can also be put in terms of our QICZ gates (Figure \ref{teleportation3}). If the particle has natural eigenstates $\ket{+}$ and $\ket{-}$ that can be easily prepared and measured, all the other operations on the particle, except for the QICZ gate, can be avoided.

Notice that the memory maps $\ket{0}$ into $\ket{+}$ and $\ket{1}$ into $\ket{-}$, as the transfer is for $H\ket{\psi}$. The reading stage introduces a second H operation that cancels the first one, recovering the original state. In fact, the transfer is incomplete until the $a$ controlled Z operation is applied. For a general qubit state $\ket{\psi}=\alpha\ket{0}+\beta\ket{1}$, the memory content is $\ket{\psi}=\alpha\ket{+}+(-1)^a\beta\ket{-}$.

Interestingly enough, we can also implement a \emph{inverting memory} if we start from an ancillary $\m$ state instead of starting with $\+$. In that case, the memory content will be $HX\ket{\psi}$ instead of $H\ket{\psi}$ and we will read a negative version of the original qubit, where the values are inverted. This X memory property will be useful for the design of the QI CNOT gates. 

If we have quantum gates that perform operations between different physical implementations of qubits, these memory schemes can transfer the information from a more volatile form into a more stable qubit, maybe less interacting and less useful for quantum gates, or less mobile and therefore not valid for quantum communication. In our proposal, optical qubits could be stored into particles and, after some time in the memory, they can be transferred back into a flying form.    

In most of the proposals for a quantum memory, a flying qubit is converted into a more stable form in another physical system. Whenever such a conversion happens, the complete write/read cycle can be seen as a teleportation, or a quantum swap, assisted by an intermediate system of a different nature. The state is transferred between qubits that are implemented on different systems. Recognizing each element of the teleportation process can help to identify the essential steps and to design simpler procedures that give the same results. 

\section{Optical CNOT gates}
\label{3vers}
The existence of a CZ gate and a quantum memory based on QI permits to create optical CNOT gates in which the interaction with an intermediary particle can create correlations between photonic qubits. The gates are better understood describing the matter part of the scheme as an ancillary qubit. Nevertheless, only a very limited set of quantum operations are applied to the particle qubits. In the following proposals, we are not dealing as much with a hybrid quantum circuit, with matter and light qubits, as with an optical qubit gate where an ancillary medium facilitates the photon interaction. This is a common feature of many optical CNOT gates that use non-linear media to complete the optical model. While the QICZ gate is perfectly symmetric and the role of the particle and the photon could be interchanged, the weight of our proposals rests on the optical part. We intend to take advantage of the simplicity of the single qubit operations on photons and reduce as much as possible the requisites on the particle part. State preparation and measurement in the $\Hbas$ basis and moderate coherence times are all that is needed for the bomb in most of the presented CNOT gates.

This section describes three families of CNOT gates. The first one is based on gate teleportation and uses the previously presented memory. The second family comes from the realization that both qubits need not to be teleported. The third family takes the reduction one step further with a configuration where there is no complete state transfer to the ancillary particle.

All the gates within each family share the same global properties, but there is a choice on the kind of classically controlled corrections and the distribution of the gates. These results are not restricted to QI gates, but can also be employed to use any CZ gate between two different kinds of qubits to give a CNOT gate between qubits of the same nature. The methods we have used to derive these gates are based on a study of equivalent quantum circuits and will be published elsewhere.
 
\subsection{Memory CNOT gate}
\label{memoryCNOT}
\begin{figure}[ht!]
\centering
\includegraphics{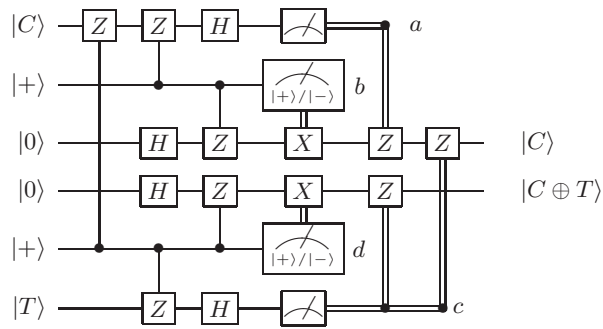}
\caption{CNOT gate through gate teleportation.}
\label{memCNOT}
\end{figure}

In a previous work \cite{GC06a}, we have shown that a quantum shutter based on QI can create the necessary entanglement to teleport the state of a photon into a particle and how a gate teleportation \cite{GC99} scheme permits to read the CNOT of the original input. The same results can be obtained with a simpler system that employs the QICZ gate presented in the preceding sections. Figure \ref{memCNOT} shows a direct translation of the quantum shutter circuit that implements an optical-to-optical CNOT gate. In the circuit, only QICZ gates, photonic H gates and measurements are needed.

In the following, $\ket{C}$ will denote the state of the control qubit of a CNOT gate and $\ket{T}$ the target state. $\ket{C}$ and $\ket{T}$ are considered to be either $\ket{0}$ or $\ket{1}$, so that $\ket{C\oplus T}$ is a XOR, or modulo 2 addition, in the usual sense. Everything is still valid for arbitrary input states, which are superpositions of the considered orthogonal basis states. In that case, $\ket{C\oplus T}$ must be understood as the input state after a CNOT operation, i.e. as a linear combination of the computational basis states where the probability amplitudes that were associated to the $\ket{C}\ket{T}$ states of the input now accompany the $\ket{C}\ket{C\oplus T}$ terms.

The teleportation steps for both control and target qubits can be understood as memory write/read cycles performed by the memory circuit of Figure \ref{teleportation3}. One possible interpretation of the CNOT operation is that of a memory cell that can be transformed into an inverting memory when $\ket{C}=\1$. As there is residual entanglement between the second memory and the control qubit, an additional sign correction is needed (in the form of a classically controlled Z gate). The undesired sign shift for input $\ket{C}=\1$ is associated to the $\1$ state of the second particle, in exactly the same way that there is an association between the sign of $\ket{T}$ and the state of the same particle (both come from a CZ operation). The sign shift that might come from measurement, when it happens, is of the same nature in both cases and the same kind of correction can be applied. This correction takes the form of the additional cZ gate of the circuit.

We can simplify the circuit by removing one of the two cZ gates of the control output line. Unless one and only one of the classical bits $a$ and $c$ are 1, the gates will do nothing or cancel each other. We can then replace the pair of gates by a single cZ gate controlled by the classical bit $a\oplus c$ with the same results. This bit can be given by a classical, and more reliable, XOR gate. 

In the proposed circuit, there are no photon-to-photon CNOT gates and, whenever possible, classical or classically controlled gates take the place of fully quantum gates. As a rule, it is worthy to avoid the quantum gates. The quantum subsystem is the most delicate part of the whole setup. Taking a part of the operation into the classical realm allows us to relax the constraints on the physical system and will lead to an easier implementation.  

This gate can be further simplified if only one of the qubits goes through a full teleportation. Next section shows a possible implementation of a half-memory CNOT gate. Nevertheless, teleporting the optical qubits at every CNOT operation can have some advantages on its own. Teleportation provides an error detector. If there is a photon loss, the teleportation fails. If no photon is found in the optical measurement, we can deduce the photon was lost and the process has to start again. Without the teleportation step, either we have to wait until the computation is supposed to finish, and loose all the intervening time, or we need to confirm, nondestructively, the presence of the photon. We will come back to the advantages of an intermediate teleportation step when all the proposed gates are compared.

Finally, it is worth remarking that we have again a decidedly non-classical effect. The CNOT interaction happens between two photons that are not in contact and it is mediated by a particle that never coincides with them in any classical sense.

\subsection{Half-memory CNOT gate}
\label{halfmemory}
From the analysis of the memory CNOT gate, it becomes obvious that the teleportation of the control qubit is not necessary. In contrast to usual gate teleportation, where entangling operations are to be avoided, here we just need to guarantee that all the CZ gates are applied between a particle and a photonic qubit. 

\begin{figure}[ht!]
\centering
\includegraphics{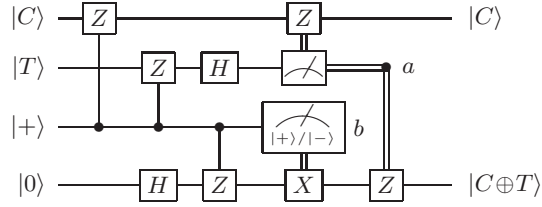}
\caption{Half-memory CNOT gate. Only one memory circuit is needed.}
\label{halfmemCNOT}
\end{figure}

We can directly take the circuit of Figure \ref{memCNOT} and suppress the teleportation/memory stage of one of the qubits. The sign correction given by the end cZ gate is still necessary. The gates can then be reordered to give the circuit of Figure \ref{halfmemCNOT}.

We can understand the gate operation as a controlled inverting memory stage, where the initial particle state is $\+$ or $\m$ depending on the value of $\ket{C}$. We can also move the CZ gate after the write step and see the process as a negation in the $\Hbas$ basis. The CNOT happens then on the stored value. Once again, $\ket{C}$ and $\ket{T}$ are entangled to the particle in the same way and need the same sign correction. Now, no qubit has two classically controlled gates of the same kind and further substitutions with classical gates, like the XOR simplification of the previous section, are not possible.

\begin{figure}[ht!]
\centering
\includegraphics{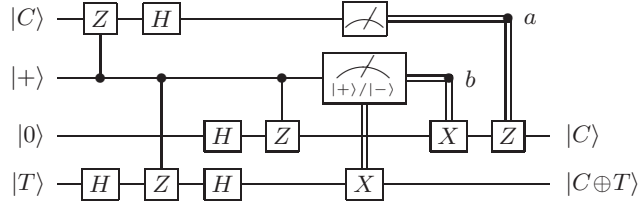}
\caption{Half-memory CNOT gate where the target qubit is kept.}
\label{halfmemtarget}
\end{figure}

The memory CNOT circuit is highly symmetrical. We can also design circuits where it is the control qubit which is teleported. Figure \ref{halfmemtarget} depicts a half-memory CNOT gate with a teleported control. The arrangement of the gates allows for an easy interpretation in terms of memory. First, the control qubit is transferred to the particle (write stage). The contents of the memory control a CNOT on the target. Later the control qubit is transferred into a new photon (read stage). The CNOT is controlled by a qubit in the $\Hbas$ basis. To correct this, depending on the value of the measurement on the particle, an X gate is applied, for $\m$, or not, if the measurement found $\+$.  

The alternative half-memory circuits offer an interesting trade-off between cZ and cX gates that depends on whether the teleported qubit is the target or the control qubit. We can alternate both CNOT types so that all the qubits are checked for photon loss every certain number of operations or use the equivalences to replace cZ gates with cX gates or vice versa, depending on the efficiencies or availability of each.

The given half-memory CNOT gates refine the previous scheme and suppress the need for a second teleportation. Still, in the whole operation control and target qubits need only one interaction with the particle. The additional CZ gate seems to appear only from the teleportation step. Without teleportation, even the memory step can be skipped.  

\subsection{Direct CNOT gate}
\label{direct}
In implementations where an efficient H gate is available in the particle system, we can also implement the CNOT gate without measuring any of the optical qubits. In this case, we can no longer hold a memory interpretation. The no cloning theorem implies that a truly quantum memory needs the destruction or the erasure of the original qubit. Without a complete transfer, part of the information or correlations that can affect the stored contents will be outside the memory. Here, the intermediate system does not act as a memory, but as a facilitator of the interaction. The concept was already employed for the control qubit of the half-memory gate. The direct CNOT gate takes this idea one step further at the cost of an extra operation on the intermediate qubit. 

\begin{figure}[ht!]
\centering
\includegraphics{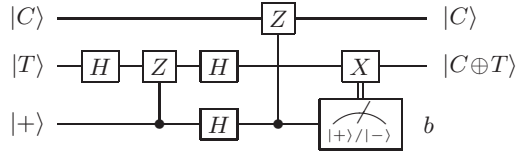}
\caption{Direct CNOT gate.}
\label{directCNOT}
\end{figure}

Figure \ref{directCNOT} shows the resulting circuit. The general complexity of the system has been reduced, but, in exchange, we need an H gate in the ancillary particle qubit.

\begin{figure}[ht!]
\centering
\includegraphics{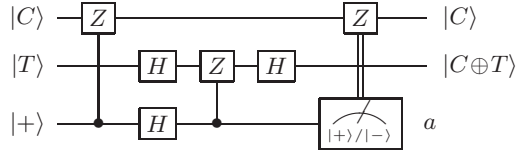}
\caption{Direct CNOT gate with cZ postcorrection.}
\label{directCNOTZ}
\end{figure}

The postcorrection can also be made through a cZ instead of a cX gate (Figure \ref{directCNOTZ}).

While these gates are the most economic in terms of CZ gates, no regeneration of the photon takes place and photon losses can happen unnoticed.

\subsection{Efficiency analysis}
\label{efficiency}
The proposed implementations offer a wide range of alternative CNOT gates to choose from. Each gate has its own set of advantages and disadvantages. The election will depend on which parameters are most important to optimize. 

The intermediate memory step, although more costly in terms of elementary gates, has some desirable properties. In long distance quantum communication, the degradation of the qubits during the transmission limits the maximum achievable distance. The probabilities of absorption and depolarization associated with propagation through optical fibre grow exponentially with the covered distance. In order to increase the communication range, the total path can be divided into subsections with the help of quantum repeaters \cite{BDC98}. There are also less resource demanding schemes that can be used with optical qubits. In particular, quantum relays offer alternatives to divide the total path into subsections that preserve better the fragile qubit state \cite{JPF02}. 

In the CNOT gates with a memory step, the teleportation of the photonic qubits provides a restricted kind of regeneration and effectively divides the total computation into subsections. The global probability of photon loss is not affected. For an attenuation coefficient of $\alpha$, the net effect of having an $e^{-\alpha L}$ absorption probability is equivalent to the effect of $n$ sections with probability of absorption  $e^{-\frac{\alpha L}{n} }$ each. The existence of teleportation steps, however, can help to detect photon losses and to avoid delay between attempts. Path subdivision can also be complemented with error correction techniques to reduce photon loss probability with the aid of quantum transponders \cite{GKL03}. Sectioning can also increase fidelity. In that case, the fidelity of the imperfect operations involved in the teleportation must be below a certain threshold that depends on the effect of the channel. A detailed analysis of the benefits of segmentation in signal to noise ratios and fidelity can be found in the literature \cite{BDC98,JPF02,DMT04}.

In the short range, where propagation losses are not so important, imperfections can also produce malfunctions like depolarization or photon loss and memory CNOT gates can still be beneficial. Memory CNOT gates seem advisable for CNOT operations in remote nodes of a quantum network. Inside quantum computers, where degradation is likely to be smaller, half-memory CNOTs can be alternated so that in every qubit there is a memory step every given number of operations. This provides an additional error check that can save time in feed-forward schemes.  

There is a further advantage in the memory and half-memory CNOT gates. In both cases the CNOT step can be implemented from QI memory blocks  and a QICZ gate. Standardization of elements is an important step towards scalability. If only a few blocks are needed, they can be produced with better quality and at a lower cost. We can even use the same block as a two qubit memory and as a CNOT gate if the first QICZ gate, which activates the inverting memory, can be externally enabled and disabled.

The direct CNOT scheme is more economical in terms of gates. The number of interactions with the particle, the most delicate part, is reduced to only two QICZ gates. In principle, having one less gate simplifies the operation. Nevertheless, the intermediate particle H operation forbids a parallelization in the QICZ operations using a setup with coupled interferometers, which is in principle possible with the other two CNOT gate models.

In terms of operation time, this will mean that, contrary to expectations, the direct CNOT gate will present a higher delay. If we consider that measurement and postcorrection together take only one gate operation time, we can arrange the gates in all of the given CNOT proposals so that there are four stages. The circuits are chosen from the variations of the basic scheme that allow the maximum number of parallel operations. In the memory CNOT gate, the number of steps is reduced with a XOR simplification. In the direct CNOT, we take the circuit with a cZ gate in the postcorrection. For this configuration, we can measure the particle and perform the last photonic H gate at the same time in order to save one stage. Notice, however, that photonic H operations are implemented with just a beamsplitter. In the total account, the $N$ cycles of the QICZ gates are the limiting factor, with the possible exception of the measurement and postcorrection steps, which are common to all the proposals. The particle H gate prevents the combination of all the QICZ operations into a single step and doubles the most important contribution to the total operation time. 

The gate number reduction is more significant when faulty gates are considered. As a rule, an increase in the number of elements introduces more factors that can go wrong, like misalignment of the elements or cumulative errors.

In a first approximation, we can assign a probability of success $p$ to the optical H gates, $q$ to the QICZ gates, $r$ to the classically controlled gates, be them cX or cZ, and $s$ to the particle H gate, and assume that particle measurement and state preparation are perfect. Optical measurement will depend on the photodetector efficiency $\eta$. A quick inspection on the proposed gates returns the count of gates and probabilities shown in Table \ref{probabilities}.

\begin{table}[ht]
\begin{center}
\begin{tabular}{@{\extracolsep{-0.24ex}}|c|c|c|c|c|c|c|}
\hline    
\phantom{\Big(}Gate& \scriptsize{H}       &          & \tiny{Classically} & \scriptsize{H}           & &\tiny{Probability} \vspace{-3.5ex}\\
         &         &  \scriptsize{QICZ}    & &              & \scriptsize{Detectors}& \vspace{-2.5ex}          \\
\phantom{\Big(}count&\scriptsize{optical}  &          &\tiny{controlled}   & \scriptsize{particle}& &\tiny{of success} \\
\hline
\scriptsize{Memory}& & & &&&\phantom{\Big(} \vspace{-2.5ex}\\
& 4& 5& 4 &0&2& $\eta^2 p^4 q^5 r^4$\phantom{\Big(} \vspace{-2.5ex}\\
\scriptsize{CNOT}& & &  &&& \phantom{\Big(}\\
\hline
\scriptsize{Half-memory}& & & &&&\phantom{\Big(} \vspace{-2.5ex}\\
& 2& 3 & 3&0 &1 & $\eta p^2 q^3 r^3$\phantom{\Big(} \vspace{-2.5ex}\\
\scriptsize{CNOT}& & &  &&& \phantom{\Big(}\\
\hline
\scriptsize{Direct}& & & &&&\phantom{\Big(} \vspace{-2.5ex}\\
& 2& 2 & 1 &1& 0& $p^2 q^2 r s$\phantom{\Big(} \vspace{-2.5ex}\\
\scriptsize{CNOT}& & &  &&& \phantom{\Big(}\\
\hline
\end{tabular}
\caption{Maximum probability of success of the different CNOT proposals when the elements are not perfect.}
\label{probabilities}
\end{center}
\end{table}

There is a minimum number of gates that are present in all the cases and we can separate a common factor of $p^2 q^2 r$, which bounds the maximum attainable probability of success for all the CNOT families. If we compare then the half-memory and direct CNOT gates, we find that direct CNOT gates have a higher probability of success if $s> \eta q r^2$. The memory CNOT gate has the same elements as the half-memory CNOT with an additional teleportation. Therefore, its probability of success can never be greater than the half-memory success rate. 

The model can be progressively refined to take into account the dependence of $q$ with the number of cycles, the particle measurement errors, or to differentiate the probabilities of error of cX and cZ. Particle measurement errors and the possible need for a pair of particle H gates at the beginning and at the end of the process if state generation and measurement in the $\Hbas$ basis are not available do not affect the probability advantage of the direct CNOT gate. In those cases, all the other CNOTs need to add at least the same correction terms. 

Under realistic conditions, $p$ can be taken to be 1. Errors in optical H gates are due to imperfections in the beamsplitter. Even in the case of faulty beamsplitters, the effect of a single photon splitting would be small in comparison with the cumulative effect of imperfections in the PBSs through the $N$ cycles of the QI step, which would make $q$ a much more important cause of failure. For a perfect operation, $q$ tends to one in the high $N$ limit. However, optical losses can seriously impair the maximum probability of success of QI \cite{KWM99}. Postcorrection errors and low detector efficiency can also be a considerable source of error \cite{KMN07}. Postcorrection is usually performed with lossy elements, like electro-optic modulators, that can make $r$ an important factor. Detector efficiency, $\eta$, is also a great concern in all LOCQ models, where false detections, or dark counts, are as important as missed photons. On top of that, particle qubits are well developed (see, for instance, \cite{CZ04}). Consequently, the condition $s> \eta q r^2$ is likely to be true, even for very efficient QI. 

For a complete comparison, it would also be interesting to have a full fidelity analysis and experimental data on the attainable probabilities for each component.

\section{Implementation}
\label{implement}
For an experimental construction of the proposed memory and CNOT gates, we need to implement two groups of operations: the optical single-qubit gates, which can be built from standard optical elements, and the QICZ gate, which depends on finding a suitable quantum system to act as the bomb. Once the quantum object is introduced into the QI scheme, the derivation of a particle controlled Z gate for dual-rail qubits is straightforward. 

The model can also be extended to polarization encoded qubits. A PBS followed by a $-\frac{\pi}{2}$ rotator in the vertical polarization branch can convert polarization encoding to dual-rail. The opposite configuration restores the polarization encoding. 

Up to this point, we have refrained from giving a concrete particle system, as many options seem feasible. But, although, in the literature, the quantum interrogation of quantum objects is deemed as plausible, no explicit experiment has been realized yet. In this section, we will draw the attention to some of the advantages of the QI gates which can be used to reduce the experimental complexity of matter to light coupling for quantum information processing. Then, we will propose a model for the realization of the bomb system using collective atomic states. 

\subsection{Complexity reduction}
The proposed gates can reach high success rates even with relatively simple setups. There are different methods to reduce the experimental requirements of the scheme. 

QI schemes have two important advantages when compared to other matter and light quantum information protocols. First, QI schemes can produce a controlled evolution even for less than perfect couplings. Second, the absorption, or scattering, mechanism needs not to be carefully designed to avoid spontaneous emission or to provide a coherent evolution of the bomb state. Absorption only happens in case of failure and the evolution beyond that point is not important. The scheme will be valid even if absorption is replaced by a decoherent process. It would equally frustrate the interference that makes up the alternative evolution, while happening with the same low probability. We can use these facts to recast some experimentally demanding quantum computer proposals into a lighter form. 

A QI gate can be simpler than other memory and gate proposals. Adding a quantum object is not as restrictive as it may seem. The coupling constraints can be relaxed if partially absorbing objects are allowed. A partially absorbing particle has a certain probability lower than 1 of absorbing a passing photon. In fact, this is the case for any quantum bomb. Interaction between a single atom and a single photon demands a strong interaction and, even for the best attainable coupling, the efficiency of the interaction is smaller than 1. 

There are various analyses of QI of partially absorbing objects under different points of view \cite{Jan99}. They all agree in the prediction of an asymptotic reproduction of the large $N$ behaviour of the perfect absorbing system when the alternative case, the no bomb case, is perfectly transmissive.

In partially absorbing bombs, however, the tendency to a perfect efficiency is slower and to obtain the same probability of success it is necessary to increase the number of cycles. In a real world experiment, where losses limit the number of cycles, the resulting probability of success can decrease, but the ability to perform simpler experiments that can be repeated at a lower cost can compensate for this disadvantage. The probability of absorption can also be boosted by using many particles in an entangled state, such as the collective atomic states of Section \ref{membomb}.

Exact fidelity results for QI systems with moderately coupled bombs should be further investigated. However, there is a qualitative argument that explains why both this scheme and QI of partially absorbing particles should present high fidelity, similar to the ideal case. If the probability of absorption is much greater than the $\sin^2(\theta)$ probability of a photon going to the upper part of the interferometer, deviations from the $\ket{H}$ state will be small in each cycle and, within a reasonable number of cycles, there will be a measurement of no $\ket{V}$ photons or an absorption will occur. At that point the procedure will be perfectly restored. The probability of error will increase because the $\ket{V}$ part of the superposition has had a few cycles to build up, but, for a sufficiently small $\theta$ and a high enough probability of absorption, this will constitute no serious hindrance to the protocol. An uncorrected residual $\ket{V}$ state will be left at the end of the $N$ cycles, when there has been no time to correct the last few absorption failures, but we can see it will have a small effect on fidelity for the same reasons. 

The choice of the $\Hbas$ basis to reduce the total number of operations is also less restrictive than it appears to be. Some atoms present polarization selective absorption. Light in different polarization states can prepare the desired atomic superpositions and be later used for the measurement. Furthermore, for some systems, the eigenstates can be $\+$, $\m$ superpositions of absorbing and non-absorbing states. Nevertheless, the restriction can be ignored if there are efficient H gates for the particle system. 

An additional advantage is that all these optical CNOT gates can be employed even for small coherence times that would render the memory unusable. For the CNOT operation, the required particle qubit lifetimes are much smaller than those of the optical qubits. They only need to live up to the $N$ cycles and the measurement step.

All these factors contribute to relax the constraints on an experimental construction and facilitate the creation of scalable optical quantum computers. Recent advances in the microelectronic integration of photon and particle manipulation tools \cite{LDL06} should allow the scaling to a higher number of qubits.

The price for using QI models is the need for optical setups with very low photon loss. QI schemes are extremely sensitive to losses, which limit the maximum number of cycles for which QI can succeed beyond a certain probability \cite{Rud00}. This also limits the correcting power against low couplings. As a whole, in QI gates, the complexity of the matter-to-light coupling is traded against a better, almost lossless optical scheme. However, reducing photon loss is an important objective in many areas of optical communications. Advances in those fields can be applied to QI gates and, even if the gates are not good enough, the efforts put into reducing losses will have a quick application into other domains.

\subsection{Implementation proposals for the bomb}
\label{ImpProposals}
The complete QI scheme will be determined by the type of the bomb. We have argued that QI techniques can be applied to simplify quantum computation implementations. In this section, we will offer concrete models as examples. The quantum bomb will be implemented using three-level atoms, either alone or forming an atomic ensemble.

The quantum bomb explosion is taken to be a photon absorption, which is associated to a certain polarization. The concrete polarization for which absorption is possible varies with the different atomic systems and their allowed level transitions. In our discussion, however, we assume that in the bomb arm the polarization of the photon is converted to the necessary state for the coupling and then converted back to vertical before reentering the interferometer. More efficient setups can be designed for each concrete case, but the principle behind the overall scheme would be the same.

\subsubsection{Single atoms inside cavities and in free-space}
\begin{figure}[ht!]
\centering
\includegraphics{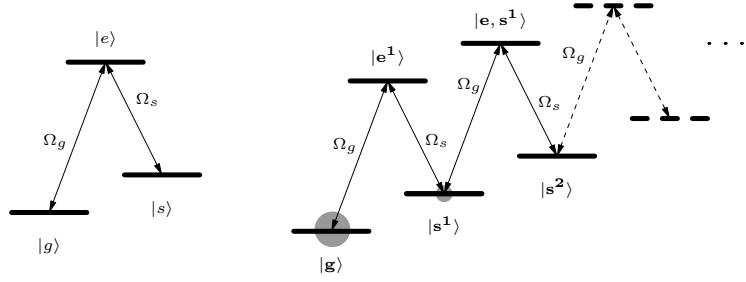}
\caption{Three level atom models for the quantum bomb with single atoms (left) and with collective atomic states in ensembles (right).}
\label{threelevelbombs}
\end{figure}

We will first consider single atoms as the bomb. The main challenge is finding a strong enough coupling between a single atom and a single photon. One way to achieve this coupling is placing the atom inside a cavity. An option that has been considered in previous QI proposals \cite{GC06a,Pav07,HM08} is a three-level atom, such as the one shown in Figure \ref{threelevelbombs} (left). The atom has two metastable states $\ket{g}$ and $\ket{s}$ and an excited state $\ket{e}$. The $\g$ and $\s$ states could be, for instance, the spin sublevels of the electronic ground state of an alkali atom. 

The photon's frequency is resonant to the $\ket{g}\rightarrow\ket{e}$ transition. An atom in $\ket{g}$ will absorb a passing photon, while an atom in $\ket{s}$ will not see the photon, which will pass unimpeded. In this system, $\g$ corresponds to the closed interferometer, or $\0$ bomb state, and $\s$ to an open interferometer, or $\1$ state. 

In the proposed QI scheme, we just need to be able to prepare and read states in the $\+=\frac{\g+\s}{\sqrt{2}}$ and $\m=\frac{\g-\s}{\sqrt{2}}$ basis, which, for single atoms, can be done with well-developed techniques. Likewise, bomb explosion could be identified, if desired, observing the occupation of the excited level $\e$ with fluorescence detection. A detailed example of these steps for the case of a Rb atom can be found in \cite{Pav07}.

A three-level atom is used to avoid some of the troubles of a system with only two levels. The excited state would also be transparent to the photon but, apart from being susceptible to spontaneous decay, the same strong coupling which makes absorption efficient, would also give a high probability of stimulated emission every time the interrogating photon goes through the open interferometer. Instead, we can use the two stable states $\g$ and $\s$. For similar reasons, three-level atoms are used in different quantum memory schemes, as we will see in the next section. 

While a cavity interaction suits well the experimental implementations of QI inside cavities \cite{Tse98}, coupling a photon in and out the cavity at each round would be difficult to realize in our scheme. We will try to find a less demanding model. It is worth remarking that, even in the cavity coupling case, less tight a coupling might suffice. QI can compensate for a lower absorption if there is a higher number of rounds. 

There are also free-space alternatives for the coupling. One interesting possibility is the shaping of the photon's wavefront so that it matches as closely as possible the time reverse of the profile of a spontaneously emitted photon. In particular, there are recent interesting theoretical and experimental results for the coupling of photons to atomic dipoles \cite{SMK07} which, despite the high losses of the preliminary experiments, could constitute an alternative to cavities. The coupling is not restricted to absorption, but also includes enhanced reflection and scattering, which, for our system, would work equally well for the bomb. 

\subsubsection{Quantum bombs with collective media}
\label{membomb}
Coupling to atomic ensembles is another option for free-space configurations. We can use the existing experimental quantum memory schemes for light as the bomb. Most of the experiments on quantum memories rely on transferring the state of a light field to a collective atomic state. Many of these approaches, such as photon-echo, Raman, and electromagnetically induced transparency quantum memories, are essentially equivalent and have an efficiency that depends only on the optical depth of a dense medium \cite{GAF07}.

Quantum memories work in three stages. In the first one, a control pulse directs the storage of a photon. The ensemble starts in a collective ground state that can absorb or be otherwise affected by a passing photon. Under the write pulse illumination, the photon is lost or reemitted into the background, but its state is mapped into the ensemble. Then, there comes a storage phase in which the state of the ensemble must be preserved. Finally, a read pulse is sent to the ensemble to stimulate a coherent emission of the stored photon. 

In a QI setup, the full memory cycle is not needed. If we can have a high absorption rate, a later incoherent emission of the photon will be acceptable for QI. Emission would only be important in the case of coordinated reemission back into the interferometer. For spontaneous emission, the probability of this event would be negligible. The photon needs not to be read in a coherent way. In the design of memory protocols, the write and read pulses have to be carefully designed in order to optimize the efficiency of the memory \cite{NGP07}. With QI, only half of the effort is required, as only the write pulse is important. 

These quantum memory schemes overcome the low coupling between light and atoms by using atomic ensembles with a collective quantum state. Collective states of different kinds have been described since the 1950s \footnote{Curiously enough, one of the first described ensemble states was proposed for what was called an ``optical bomb'', although for completely different reasons. Dicke explains how this name appeared in an early model of a laser where many atoms in a collective state emitted a burst of light in a coherent and coordinated way \cite{Dic64}. Interestingly, Dicke did not only give the first descriptions of collective states \cite{Dic54}, but also was among the first to point out the importance of negative measurements and counterfactuals in quantum theory \cite{Dic81}, from which later sprang the ideas of quantum interrogation.}. In the quantum memories, the collective state of the atoms still show quantum behaviour while presenting, as a joint system, a size of many optical wavelengths. The interaction is efficient, as any of the atoms of the ensemble can absorb the photon with the same result, increasing the absorption rate in a way proportional to their number.

We will use the quantum memory of \cite{FL02}, but similar models can be devised for other kinds of quantum memories. We consider an ensemble of $N_a$ three-level atoms. All the atoms in the ensemble can be prepared in the $\g$ state using standard techniques. We will call this collective ground state $\ket{\mathbf{g}}\equiv\ket{g_1}\cdots\ket{g_{N_a}}$, where $\ket{g_i}$ is the ground state of the $i$th atom of the ensemble. With a signal field $\Omega_g$ and a control field $\Omega_s$, we can produce collective symmetric states $\ket{\mathbf{s^1}}=\frac{1}{\sqrt{N_a}}\sum_{i=1}^{N_a}\ket{g_1}\cdots\ket{s_i}\cdots\ket{g_{N_a}}$ with a superposition of terms where one, and only one, of the atoms has been excited and then taken to the stable state $\s$. The photon is now stored into the $\sbold$ collective state.

Under the particular conditions the atomic sample, it is correct to speak of only one $\ket{\mathbf{s^1}}$ state, as the atoms cannot be distinguished. In the case of $N_a$ separated atoms, the photon absorption can be rightfully said to have occurred at one identifiable atom. For this ensemble, however, the distinction disappears and an absorption will always produce the same collective state. We can, likewise, define an $\ket{\mathbf{s^n}}$ state where $n$ photons have been absorbed and transfered to an $\s$ state. The ladder of transitions is shown on the right of Figure \ref{threelevelbombs}. 

For a small enough value of $n$, both $\ket{\mathbf{g}}$ and $\ket{\mathbf{s^n}}$ states show an enhanced coupling to light and have high probability of absorbing single photons. Unlike in the quantum memory proposals, we will not be concerned with storing a photon state, but we will try to have superpositions of a highly absorptive state and a transparent state. An ensemble in state $\ket{\mathbf{g}}$ can act as an active bomb. If we transfer all the atoms to an $\s$ state and have an $\ket{\mathbf{s^{N_a}}}$ state, light will not be absorbed. Small errors can be tolerated, as states with few atoms in the $\g$ state will not show the enhanced coupling to light of $\ket{\mathbf{g}}$. 

This direct approach could be used with different quantum memory schemes. We can imagine we store $N_a$ photons one by one or many at a time. The resulting cat-like states $\frac{1}{\sqrt{2}}[\gb+\sbN]$ have already been proposed for improved metrology \cite{BIW96} and do not necessarily require a quantum memory for their production. However, while states like $\sbold$ are resilient to decoherence and quite robust despite being highly entangled, the cat superpositions $\frac{1}{\sqrt{2}}[\gb+\sbN]$ are more delicate \cite{DVC00}. It would be desirable to find a different transparent state for which the superpositions would be more stable.

A good option are the memory proposals with Rydberg atoms that show dipole blockade. We will sketch how the QI scheme can work for an atomic ensemble in a cold Rydberg gas, such as the one described in \cite{LFC01}.

The states relevant to our proposal are the states $\g$ and $\s$ of the ground state manifold and long-lived Rydberg states $\ket{r}$, $\ket{p'}$ and $\ket{p''}$. For small atomic densities, the interactions between the ground states are negligible, but Rydberg states are strongly coupled due to dipole-dipole interactions. One important property of this ensemble is that the dipole-dipole interactions will strongly inhibit many of the in principle possible transitions. In our case, the dipole blockade will guarantee that only the $\ket{r}$ Rydberg state is excited and that collective states with more that one excitation are highly unlikely. The resulting level system is similar to the three-level atom model where $\ket{r}$ has replaced $\e$ and it is impossible to have more than a single excitation at the same time. 

We can imagine that we have two states, $\ket{\mathbf{g}}$, for which the absorption of light tuned to the $\ket{g}\rightarrow\ket{r}$ transition is allowed and a $\ket{\mathbf{r^1}}$ state, for which no further absorption is possible for light resonant to the same transition. By adequate control fields, these ensembles can be prepared in arbitrary superpositions of collective states $\alpha_0\ket{\mathbf{g}}+\alpha_1\ket{\mathbf{\mathbf{r^1}}}$, with $|\alpha_0|^2+|\alpha_1|^2=1$ \cite{LFC01}. In these states, the ensemble acts as a two-level atom in a superposition of being in the ground and the excited states. These superpositions are stable. Rydberg states are long lived and spontaneous emission should be negligible during the QI procedure. 

For a photon resonant to the $\ket{g}\rightarrow\ket{r}$ transition, either by itself or with the help of a background field in a two-photon process, the state $\ket{\mathbf{r^1}}$ is transparent and the $\ket{\mathbf{g}}$ state has $N_a$ atoms that can absorb the passing photon. This will result in an improved coupling. In contrast, the spontaneous emission rate from $\ket{\mathbf{r^1}}$ will be small, as only one atom is implied.

Even though the coherence time for the $\ket{\mathbf{r^1}}$ is high, it is not valid as the transparent state. The high coupling between the photon and the $\gb$ and the $\ket{\mathbf{r^1}}$ states, which is necessary for a high absorption, also implies a high rate of stimulated emission. The observed Rabi oscillations between the $\gb$ and $\ket{\mathbf{r^1}}$ states show this in a clear manner. Unlike in the case of spontaneous emission, stimulated emission is a collective process and will be coherently enhanced. This makes $\ket{\mathbf{r^1}}$ a poor transparent state, not because it will absorb the interrogating photon, which it will not, but because of the risk of stimulating an emission from $\ket{\mathbf{r^1}}$. That emission would destroy the state of the quantum bomb and make the QI protocol fail.

We will propose a variation on the blockade theme related to the dipole blockade CNOT gate between a single Rydberg atom and an atomic ensemble of \cite{MLW08}. To create the initial bomb state, we prepare the individual Rydberg atom in a $\frac{\ket{g}+\ket{r}}{\sqrt{2}}$ superposition, for instance by shining on the atom a laser pulse of the appropriate duration. Then, the atom is moved close to an ensemble in the $\gb$ state at a distance that allows the long range dipole blockade to show. At this point we have the superposition $\frac{1}{\sqrt{2}}[\g\gb+\ket{r}\gb]$. The state $\g\gb$ shows an enhanced absorption, but the dipole blockade prevents any absorption from the $\ket{r}\gb$ state. Notice, however, that now stimulated emission is not collective. Both spontaneous and stimulated emission depend on a single atom. Moreover, the excited atom can even be outside the path of the photon, which only needs to cross the ensemble. Absorption, if it happens, would result in a collective state that could be detected. When compared to the proposal of \cite{MLW08}, this QI model has to its advantage that we do not need either arbitrary state preparation for the single atom or the generation of complex cat ensemble states.

To complete the QI protocol, we just need to take back the single atom and check whether its state is $\frac{\g+\ket{r}}{\sqrt{2}}$ or $\frac{\g-\ket{r}}{\sqrt{2}}$, which can also be done with standard techniques. Initialization and readout of the single atom state is slower than the preparation and readout of the ensemble superpositions of $\gb$ and $\ket{\mathbf{r^1}}$, but we are not worried about long storage times and will only need to perform two of such operations. If the readout time is still too long, there are proposals for more efficient measurement schemes in which the atom is coupled to ensembles \cite{SW08}.

The central element of all the dipole blockade gates, including our QI proposal, is the strong suppression of multiple excitations. Up to date, experimental observation of the dipole blockade effect has been restricted to a small number of atoms, but recent results suggest that proposals based on the mesoscopic dipole blockade will be a viable alternative in the near future \cite{VVZ06}.

These examples show that quantum memories can be modified to act as the quantum bomb. We need a system with, at least, an state with high absorption and another state transparent to the photon. Additionally, we need to be able to prepare the bomb in a superposition of being active and not and to measure its state in the $\Hbas$ basis. The demands for coherence are limited to being able to maintain those highly entangled states. The difficulty to create and preserve them will depend on the available transparent states of the memory scheme at hand. 

The QI scheme will work even if the concrete atom that absorbed the photon can be identified. As soon as an absorption happens, the protocol fails. This means we do not care if the state resulting from an explosion is not collective. Even the single excitation of one of $N_a$ atoms would act as the quantum bomb explosion. The lack of a photon in the corresponding mode will indicate a protocol failure. 

Under these circumstances, QI can be used to create optical CNOT gates out of quantum memories. The success of the model depends on keeping the optical losses low. If this is achieved, the experimental realization of the QI gates can even be simpler than the construction of the usual quantum memories. One of the greatest advantages of the ensemble approach is that cavities are not needed for the coupling. In most of these memories, the interaction between the light and the ensemble happens in a free-space configuration. We can avoid coupling the light in and out the cavity, cutting an important source of losses. The overall scheme is simplified without the cavity.

The atomic ensembles can be controlled by relatively simple laser configurations and can even be created at warm temperatures. There are many successful experimental demonstrations of quantum memories that could be adapted for QI \cite{JSC04}. In fact, similar systems have already been employed in initial experiments in applications related to interaction-free measurement with good results \cite{CPA08}.

\subsection{Comparison with other optical proposals}
The QI model has two main advantages over other linear optical proposals. First, unlike many optical systems, it provides a method for coupling the information to a quantum memory with the same elements that are used for the gates. No additional elements are needed and the same standard blocks can be employed. Second, it can substantially improve the probability of success in the CNOT operation.

Experimentally demonstrated CNOT gates, even for perfect operation, cannot exceed a success probability of $\frac{1}{4}$ \cite{GPW04}. Linear optics quantum gates can improve their success rates using postselection and postcorrection, but only at the price of increasing the number of the ancillae \cite{FDF02}. To achieve efficiencies comparable to the estimations for the QI models, more complicated circuits, with more room for imperfection, would be needed. We won't discuss the proposals for improving the probabilities of postselection gates using error correction and teleportation techniques. Those procedures can also be applied to the QI case, where the greater CNOT operation efficiency will improve the overall result. 

Similarly, most the existing gate proposals with QI are either destructive, limited or probabilistic (see Section \ref{comp}). Among them, a particularly interesting model was the probabilistic optical CNOT gate based on QI of quantum objects in a photon resonator of \cite{Pav07}. We have shown that cavity coupling is not strictly necessary and how less demanding coupling mechanisms could suffice. The other deterministic optical scheme of \cite{HM08}, while having the advantage of not requiring measurement, needs more advanced control of the particle qubit. In particular, it requires the ability to perform arbitrary rotations on the particle's state.

This by no means exhaustive account of possible systems illustrates the ambit in which the QI of quantum particles model can be advantageously applied. As most of the operations can be left to the optical part, the global system can show a greater simplicity than their non-interrogation counterparts while showing a better probability of success than the existing linear optical schemes.

\section{Discussion}
\label{discussion}
We have presented different applications of quantum interrogation of quantum objects. The ability to create entanglement between photons and particles without the need of any actual absorption or scattering taking place gives new ways to produce the elusive photon interaction that is a prerequisite for optical quantum computation. This interaction is provided through a particle to photon CZ gate. 

First, we have put forward simple optical setups that can produce the complex entangled states that are a fundamental resource for many quantum applications. The photons are entangled with the help on an intermediate particle system.

Inspired by the quantum gate teleportation model, we can use the interaction with ancillary systems to create optical CNOT gates. As opposed to most of the previous proposals of entangling gates based on quantum interrogation, these CNOT gates are deterministic in the perfect operation limit and have no restriction on the input states. We have tried to simplify the physical implementation and substitute quantum gates with classical ones, less fragile and less prone to error.

Three families of CNOT gates result from this process. In memory CNOT gates, both input qubits are teleported and can be checked for errors. Half-memory CNOT gates put just one of the qubits through the memory step. Direct CNOT gates only use the particle as an ancillary system that never has the complete state of any of the qubits. Each family has its own strong and weak points. 

Quantum interrogation of quantum objects, when combined with techniques from linear optics quantum computation, provides all the elements necessary for a complete quantum information system. Both the presented CNOT gates and the QI setups for the generation of entangled optical states are enough to complete linear optics in the KLM scheme. 

The proposed quantum computation model with quantum interrogation can be interpreted as a measurement based quantum computation. The memory and half-memory CNOT gates are an example of teleportation based quantum computation, which, in turn, is equivalent to the one-way quantum computer scheme \cite{AL04} and the whole KLM model is, itself, a kind of measurement based computation \cite{Pop07}. In fact, the application of the Zeno effect in quantum interrogation can be seen as an extreme form of measurement based computation, where continuous measurement has endowed us with a nearly deterministic CNOT gate that avoids employing a large number of ancillae.

The optical elements of the scheme have already been used in linear optics quantum computation experiments. For the particle part, we need a system able to interact with the photons and keep superposition for a long enough period. Assuming such a particle, a general scheme based in quantum interrogation has been derived. The proposed systems use linear optical elements and an absorptive element (the bomb). There are several options for the bomb system. We have proposed an example bomb with collective atomic states like the ones used in quantum memory experiments. With the proposed QI model, quantum memory schemes can be converted into CNOT gates. 

For the construction of a quantum interrogation memory the bomb system would need to keep superpositions of the stored states for a long time. However, the QI logic gates could be built even for particles with short coherence times comparable to the operation time of the gate. The major challenge, as in all QI schemes, is limiting the optical losses. 
  
We have shown that a fully operational quantum computer can be built using only quantum interrogation of quantum objects and linear optics. The possible applications include the construction of quantum memories and the generation of sophisticated entangled states. All the elements needed are widely available and have been successfully used, separately, in practical systems. With the proposed schemes, quantum computers based on quantum interrogation could provide simpler and scalable alternatives for quantum information processing.

\section*{ACKNOWLEDGEMENTS}
J.C.G.E would like to thank Alessandro Cer\`{e} for information on the use of collective media in interaction-free experiments. This work has been funded by projects TEC2007-67429-C02-01 of the Spanish MICINN and VA001A08 of the JCyL.

\newcommand{\noopsort}[1]{} \newcommand{\printfirst}[2]{#1}
  \newcommand{\singleletter}[1]{#1} \newcommand{\switchargs}[2]{#2#1}


\begin{thebibliography}{91}
\expandafter\ifx\csname natexlab\endcsname\relax\def\natexlab#1{#1}\fi
\expandafter\ifx\csname bibnamefont\endcsname\relax
  \def\bibnamefont#1{#1}\fi
\expandafter\ifx\csname bibfnamefont\endcsname\relax
  \def\bibfnamefont#1{#1}\fi
\expandafter\ifx\csname citenamefont\endcsname\relax
  \def\citenamefont#1{#1}\fi
\expandafter\ifx\csname url\endcsname\relax
  \def\url#1{\texttt{#1}}\fi
\expandafter\ifx\csname urlprefix\endcsname\relax\def\urlprefix{URL }\fi
\providecommand{\bibinfo}[2]{#2}
\providecommand{\eprint}[2][]{\url{#2}}

\bibitem[{\citenamefont{Nielsen and Chuang}(2000)}]{NC00}
\bibinfo{author}{\bibfnamefont{M.~A.} \bibnamefont{Nielsen}} \bibnamefont{and}
  \bibinfo{author}{\bibfnamefont{I.~L.} \bibnamefont{Chuang}},
  \emph{\bibinfo{title}{Quantum Computation and Quantum Information}}
  (\bibinfo{publisher}{Cambridge University Press},
  \bibinfo{address}{Cambridge, UK}, \bibinfo{year}{2000}),
  \bibinfo{edition}{1st} ed.

\bibitem[{\citenamefont{Knill et~al.}(2001)\citenamefont{Knill, Laflamme, and
  Milburn}}]{KLM01}
\bibinfo{author}{\bibfnamefont{E.}~\bibnamefont{Knill}},
  \bibinfo{author}{\bibfnamefont{R.}~\bibnamefont{Laflamme}}, \bibnamefont{and}
  \bibinfo{author}{\bibfnamefont{G.}~\bibnamefont{Milburn}},
  \bibinfo{journal}{Nature} \textbf{\bibinfo{volume}{409}}, \bibinfo{pages}{46}
  (\bibinfo{year}{2001}).

\bibitem[{\citenamefont{{O'Brien} et~al.}(2003)\citenamefont{{O'Brien},
  {Pryde}, {White}, {Ralph}, and {Branning}}}]{OPW03}
\bibinfo{author}{\bibfnamefont{J.~L.} \bibnamefont{{O'Brien}}},
  \bibinfo{author}{\bibfnamefont{G.~J.} \bibnamefont{{Pryde}}},
  \bibinfo{author}{\bibfnamefont{A.~G.} \bibnamefont{{White}}},
  \bibinfo{author}{\bibfnamefont{T.~C.} \bibnamefont{{Ralph}}},
  \bibnamefont{and}
  \bibinfo{author}{\bibfnamefont{D.}~\bibnamefont{{Branning}}},
  \bibinfo{journal}{Nature} \textbf{\bibinfo{volume}{426}},
  \bibinfo{pages}{264} (\bibinfo{year}{2003}).
\bibinfo{author}{\bibfnamefont{T.~B.} \bibnamefont{Pittman}},
  \bibinfo{author}{\bibfnamefont{M.~J.} \bibnamefont{Fitch}},
  \bibinfo{author}{\bibfnamefont{B.~C.} \bibnamefont{Jacobs}},
  \bibnamefont{and} \bibinfo{author}{\bibfnamefont{J.~D.}
  \bibnamefont{Franson}}, \bibinfo{journal}{Physical Review A}
  \textbf{\bibinfo{volume}{68}}, \bibinfo{pages}{032316}
  (\bibinfo{year}{2003}). 
\bibinfo{author}{\bibfnamefont{K.}~\bibnamefont{Sanaka}},
  \bibinfo{author}{\bibfnamefont{T.}~\bibnamefont{Jennewein}},
  \bibinfo{author}{\bibfnamefont{J.-W.} \bibnamefont{Pan}},
  \bibinfo{author}{\bibfnamefont{K.}~\bibnamefont{Resch}}, \bibnamefont{and}
  \bibinfo{author}{\bibfnamefont{A.}~\bibnamefont{Zeilinger}},
  \bibinfo{journal}{Physical Review Letters} \textbf{\bibinfo{volume}{92}},
  \bibinfo{pages}{017902} (\bibinfo{year}{2004}).

\bibitem[{\citenamefont{Gottesman and Chuang}(1999)}]{GC99}
\bibinfo{author}{\bibfnamefont{D.}~\bibnamefont{Gottesman}} \bibnamefont{and}
  \bibinfo{author}{\bibfnamefont{I.~L.} \bibnamefont{Chuang}},
  \bibinfo{journal}{{Nature}} \textbf{\bibinfo{volume}{402}},
  \bibinfo{pages}{390} (\bibinfo{year}{1999}).

\bibitem[{\citenamefont{Barenco et~al.}(1995)\citenamefont{Barenco, Bennett,
  Cleve, DiVincenzo, Margolus, Shor, Sleator, Smolin, and Weinfurter}}]{BBC95}
\bibinfo{author}{\bibfnamefont{A.}~\bibnamefont{Barenco}},
  \bibinfo{author}{\bibfnamefont{C.~H.} \bibnamefont{Bennett}},
  \bibinfo{author}{\bibfnamefont{R.}~\bibnamefont{Cleve}},
  \bibinfo{author}{\bibfnamefont{D.~P.} \bibnamefont{DiVincenzo}},
  \bibinfo{author}{\bibfnamefont{N.}~\bibnamefont{Margolus}},
  \bibinfo{author}{\bibfnamefont{P.}~\bibnamefont{Shor}},
  \bibinfo{author}{\bibfnamefont{T.}~\bibnamefont{Sleator}},
  \bibinfo{author}{\bibfnamefont{J.~A.} \bibnamefont{Smolin}},
  \bibnamefont{and}
  \bibinfo{author}{\bibfnamefont{H.}~\bibnamefont{Weinfurter}},
  \bibinfo{journal}{{Physical Review A}} \textbf{\bibinfo{volume}{52}},
  \bibinfo{pages}{3457} (\bibinfo{year}{1995}).

\bibitem[{\citenamefont{Kok et~al.}(2007)\citenamefont{Kok, Munro, Nemoto,
  Ralph, Dowling, and Milburn}}]{KMN07}
\bibinfo{author}{\bibfnamefont{P.}~\bibnamefont{Kok}},
  \bibinfo{author}{\bibfnamefont{W.~J.} \bibnamefont{Munro}},
  \bibinfo{author}{\bibfnamefont{K.}~\bibnamefont{Nemoto}},
  \bibinfo{author}{\bibfnamefont{T.~C.} \bibnamefont{Ralph}},
  \bibinfo{author}{\bibfnamefont{J.~P.} \bibnamefont{Dowling}},
  \bibnamefont{and} \bibinfo{author}{\bibfnamefont{G.~J.}
  \bibnamefont{Milburn}}, \bibinfo{journal}{Reviews of Modern Physics}
  \textbf{\bibinfo{volume}{79}}, \bibinfo{eid}{135} (\bibinfo{year}{2007}).

\bibitem[{\citenamefont{Fleischhauer and Lukin}(2002)}]{FL02}
\bibinfo{author}{\bibfnamefont{M.}~\bibnamefont{Fleischhauer}}
  \bibnamefont{and} \bibinfo{author}{\bibfnamefont{M.~D.} \bibnamefont{Lukin}},
  \bibinfo{journal}{{Physical Review A}} \textbf{\bibinfo{volume}{65}},
  \bibinfo{pages}{022314} (\bibinfo{year}{2002}).

\bibitem[{\citenamefont{Julsgaard}(2004)}]{JSC04}
\bibinfo{author}{\bibfnamefont{B.}~\bibnamefont{Julsgaard}},
  \bibinfo{author}{\bibfnamefont{J.}~\bibnamefont{Sherson}},
  \bibinfo{author}{\bibfnamefont{J.~I.} \bibnamefont{Cirac}},
  \bibinfo{author}{\bibfnamefont{J.}~\bibnamefont{Fiur\'a\v{s}ek}},
  \bibnamefont{and} \bibinfo{author}{\bibfnamefont{E.~S.}
  \bibnamefont{Polzik}}, \bibinfo{journal}{{Nature}}
  \textbf{\bibinfo{volume}{432}}, \bibinfo{pages}{482} (\bibinfo{year}{2004}).
\bibinfo{author}{\bibfnamefont{T.}~\bibnamefont{Chaneli\`{e}re}},
  \bibinfo{author}{\bibfnamefont{D.~N.} \bibnamefont{Matsukevich}},
  \bibinfo{author}{\bibfnamefont{S.~D.} \bibnamefont{Jenkinsm}},
  \bibinfo{author}{\bibfnamefont{S.-Y.} \bibnamefont{Lan}},
  \bibinfo{author}{\bibfnamefont{T.~A.~B.} \bibnamefont{Kennedy}},
  \bibnamefont{and} \bibinfo{author}{\bibfnamefont{A.}~\bibnamefont{Kuzmich}},
  \bibinfo{journal}{{Nature}} \textbf{\bibinfo{volume}{438}},
  \bibinfo{pages}{833} (\bibinfo{year}{2005}).
\bibinfo{author}{\bibfnamefont{L.~M.} \bibnamefont{Duan}},
  \bibinfo{author}{\bibfnamefont{A.}~\bibnamefont{Kuzmich}}, \bibnamefont{and}
  \bibinfo{author}{\bibfnamefont{H.~J.} \bibnamefont{Kimble}},
  \bibinfo{journal}{{Physical Review A}} \textbf{\bibinfo{volume}{67}},
  \bibinfo{pages}{032305} (\bibinfo{year}{2003}).

\bibitem[{\citenamefont{Dicke}(1981)}]{Dic81}
\bibinfo{author}{\bibfnamefont{R.~H.} \bibnamefont{Dicke}},
  \bibinfo{journal}{{American Journal of Physics}}
  \textbf{\bibinfo{volume}{49}}, \bibinfo{pages}{925} (\bibinfo{year}{1981}). 


\bibitem[{\citenamefont{Vaidman}(2001)}]{Vai01}
\bibinfo{author}{\bibfnamefont{L.}~\bibnamefont{Vaidman}},
  \bibinfo{journal}{Zeitschrift f\"{u}r Naturforschung A}
  \bibinfo{pages}{100--107} (\bibinfo{year}{2001}).
\bibinfo{author}{\bibfnamefont{A.~C.} \bibnamefont{Elitzur}} \bibnamefont{and}
\bibinfo{author}{\bibfnamefont{L.}~\bibnamefont{Vaidman}},
  \bibinfo{journal}{{Foundations of Physics}}
  \textbf{\bibinfo{volume}{23}}, \bibinfo{pages}{987} (\bibinfo{year}{1993}).
\bibinfo{author}{\bibfnamefont{P.}~\bibnamefont{Kwiat}},
  \bibinfo{author}{\bibfnamefont{H.}~\bibnamefont{Weinfurter}},
  \bibinfo{author}{\bibfnamefont{T.}~\bibnamefont{Herzog}},
  \bibinfo{author}{\bibfnamefont{A.}~\bibnamefont{Zeilinger}},
  \bibnamefont{and} \bibinfo{author}{\bibfnamefont{M.~A.}
  \bibnamefont{Kasevich}}, \bibinfo{journal}{{Physical Review Letters}}
  \textbf{\bibinfo{volume}{74}}, \bibinfo{pages}{4763} (\bibinfo{year}{1995}). 


\bibitem[{\citenamefont{Tsegaye}(1998)}]{Tse98}
\bibinfo{author}{\bibfnamefont{T.}~\bibnamefont{Tsegaye}},
  \bibinfo{author}{\bibfnamefont{E.}~\bibnamefont{Goobar}},
  \bibinfo{author}{\bibfnamefont{A.}~\bibnamefont{Karlsson}},
  \bibinfo{author}{\bibfnamefont{G.}~\bibnamefont{Bj\"{o}rk}},
  \bibinfo{author}{\bibfnamefont{M.~Y.} \bibnamefont{Loh}}, \bibnamefont{and}
  \bibinfo{author}{\bibfnamefont{K.~H.} \bibnamefont{Lim}},
  \bibinfo{journal}{{Physical Review A}} \textbf{\bibinfo{volume}{57}},
  \bibinfo{pages}{3987} (\bibinfo{year}{1998}).


\bibitem[{\citenamefont{Kwiat et~al.}(1999)\citenamefont{Kwiat, White,
  Mitchell, Nairz, Weihs, Weinfurter, and Zeilinger}}]{KWM99}
\bibinfo{author}{\bibfnamefont{P.~G.}~\bibnamefont{Kwiat}},
  \bibinfo{author}{\bibfnamefont{A.~G.} \bibnamefont{White}},
  \bibinfo{author}{\bibfnamefont{J.~R.} \bibnamefont{Mitchell}},
  \bibinfo{author}{\bibfnamefont{O.}~\bibnamefont{Nairz}},
  \bibinfo{author}{\bibfnamefont{G.}~\bibnamefont{Weihs}},
  \bibinfo{author}{\bibfnamefont{H.}~\bibnamefont{Weinfurter}},
  \bibnamefont{and}
  \bibinfo{author}{\bibfnamefont{A.}~\bibnamefont{Zeilinger}},
  \bibinfo{journal}{{Physical Review Letters}}
  \textbf{\bibinfo{volume}{83}}, \bibinfo{pages}{4725} (\bibinfo{year}{1999}).

\bibitem[{\citenamefont{Gilchrist et~al.}(2002)\citenamefont{Gilchrist, White,
  and Munro}}]{GWM02}
\bibinfo{author}{\bibfnamefont{A.}~\bibnamefont{Gilchrist}},
  \bibinfo{author}{\bibfnamefont{A.~G.} \bibnamefont{White}}, \bibnamefont{and}
  \bibinfo{author}{\bibfnamefont{W.~J.} \bibnamefont{Munro}},
  \bibinfo{journal}{Physical Review A} \textbf{\bibinfo{volume}{66}},
  \bibinfo{pages}{012106} (\bibinfo{year}{2002}).

\bibitem[{\citenamefont{Azuma}(2003)}]{Azu03}
\bibinfo{author}{\bibfnamefont{H.}~\bibnamefont{Azuma}},
  \bibinfo{journal}{Physical Review A} \textbf{\bibinfo{volume}{68}},
  \bibinfo{pages}{022320} (\bibinfo{year}{2003}).

\bibitem[{\citenamefont{Azuma}(2004)}]{Azu04}
\bibinfo{author}{\bibfnamefont{H.}~\bibnamefont{Azuma}},
  \bibinfo{journal}{Physical Review A} \textbf{\bibinfo{volume}{70}},
  \bibinfo{eid}{012318} (\bibinfo{year}{2004}).

\bibitem[{\citenamefont{Pavi\v{c}i\'c}(2007)}]{Pav07}
\bibinfo{author}{\bibfnamefont{M.}~\bibnamefont{Pavi\v{c}i\'c}},
  \bibinfo{journal}{Physical Review A} \textbf{\bibinfo{volume}{75}},
  \bibinfo{eid}{032342} (\bibinfo{year}{2007}).

\bibitem[{\citenamefont{Garc\'ia-Escart\'in and Chamorro-Posada}(2006)}]{GC06a}
\bibinfo{author}{\bibfnamefont{J.~C.} \bibnamefont{Garc\'ia-Escart\'in}}
  \bibnamefont{and}
  \bibinfo{author}{\bibfnamefont{P.}~\bibnamefont{Chamorro-Posada}},
  \bibinfo{journal}{Quantum Information and Computation}
  \textbf{\bibinfo{volume}{6}}, \bibinfo{pages}{495} (\bibinfo{year}{2006}).

\bibitem[{\citenamefont{Huang and Moore}(2008)}]{HM08}
\bibinfo{author}{\bibfnamefont{Y.~P.} \bibnamefont{Huang}} \bibnamefont{and}
  \bibinfo{author}{\bibfnamefont{M.~G.} \bibnamefont{Moore}},
  \bibinfo{journal}{Physical Review A} \textbf{\bibinfo{volume}{77}},
  \bibinfo{eid}{062332} (\bibinfo{year}{2008}).

\bibitem[{\citenamefont{Wang and You}(2008)}]{WYN08}
\bibinfo{author}{\bibfnamefont{X.-B.} \bibnamefont{Wang}},
  \bibinfo{author}{\bibfnamefont{J.~Q.} \bibnamefont{You}}, \bibnamefont{and}
  \bibinfo{author}{\bibfnamefont{F.}~\bibnamefont{Nori}},
  \bibinfo{journal}{Physical Review A} \textbf{\bibinfo{volume}{77}},
  \bibinfo{eid}{062339} (\bibinfo{year}{2008}).

\bibitem[{\citenamefont{Methot and Wicker}(2001)}]{MW01}
\bibinfo{author}{\bibfnamefont{A.}~\bibnamefont{M\'ethot}} \bibnamefont{and}
  \bibinfo{author}{\bibfnamefont{K.}~\bibnamefont{Wicker}},
  \bibinfo{journal}{arXiv:quant-ph/0109105v1}  (\bibinfo{year}{2001}).


\bibitem[{\citenamefont{Franson et~al.}(2004)\citenamefont{Franson, Jacobs, and
  Pittman}}]{FJP04}
\bibinfo{author}{\bibfnamefont{J.~D.} \bibnamefont{Franson}},
  \bibinfo{author}{\bibfnamefont{B.~C.} \bibnamefont{Jacobs}},
  \bibnamefont{and} \bibinfo{author}{\bibfnamefont{T.~B.}
  \bibnamefont{Pittman}}, \bibinfo{journal}{Physical Review A}
  \textbf{\bibinfo{volume}{70}}, \bibinfo{eid}{062302}
  (\bibinfo{year}{2004}).
\bibinfo{author}{\bibfnamefont{J.}~\bibnamefont{Franson}},
  \bibinfo{author}{\bibfnamefont{T.}~\bibnamefont{Pittman}}, \bibnamefont{and}
  \bibinfo{author}{\bibfnamefont{B.}~\bibnamefont{Jacobs}},
  \bibinfo{journal}{Journal of the Optical Society of America B}
  \textbf{\bibinfo{volume}{24}}, \bibinfo{pages}{209} (\bibinfo{year}{2007}).


\bibitem[{\citenamefont{Pan et~al.}(1998)\citenamefont{Pan, Bouwmeester,
  Weinfurter, and Zeilinger}}]{PBW98}
\bibinfo{author}{\bibfnamefont{J.-W.} \bibnamefont{Pan}},
  \bibinfo{author}{\bibfnamefont{D.}~\bibnamefont{Bouwmeester}},
  \bibinfo{author}{\bibfnamefont{H.}~\bibnamefont{Weinfurter}},
  \bibnamefont{and}
  \bibinfo{author}{\bibfnamefont{A.}~\bibnamefont{Zeilinger}},
  \bibinfo{journal}{Physical Review Letters} \textbf{\bibinfo{volume}{80}},
  \bibinfo{pages}{3891} (\bibinfo{year}{1998}).


\bibitem[{\citenamefont{Pittman et~al.}(2001)\citenamefont{Pittman, Jacobs, and
  Franson}}]{PJF01}
\bibinfo{author}{\bibfnamefont{T.~B.} \bibnamefont{Pittman}},
  \bibinfo{author}{\bibfnamefont{B.~C.} \bibnamefont{Jacobs}},
  \bibnamefont{and} \bibinfo{author}{\bibfnamefont{J.~D.}
  \bibnamefont{Franson}}, \bibinfo{journal}{Physical Review A}
  \textbf{\bibinfo{volume}{64}}, \bibinfo{pages}{062311}
  (\bibinfo{year}{2001}).


\bibitem[{\citenamefont{Gasparoni et~al.}(2004)\citenamefont{Gasparoni, Pan, Walther, Rudolph, and Zeilinger}}]{GPW04}
 \bibinfo{author}{\bibfnamefont{S.}~\bibnamefont{Gasparoni}},
  \bibinfo{author}{\bibfnamefont{J.-W.} \bibnamefont{Pan}},
  \bibinfo{author}{\bibfnamefont{P.}~\bibnamefont{Walther}},
  \bibinfo{author}{\bibfnamefont{T.}~\bibnamefont{Rudolph}}, \bibnamefont{and}
  \bibinfo{author}{\bibfnamefont{A.}~\bibnamefont{Zeilinger}},
  \bibinfo{journal}{{Physical Review Letters}}
  \textbf{\bibinfo{volume}{93}}, \bibinfo{pages}{020504}
  (\bibinfo{year}{2004}).


\bibitem[{\citenamefont{D\"ur et~al.}(2000)\citenamefont{D\"ur, Vidal, and
  Cirac}}]{DVC00}
\bibinfo{author}{\bibfnamefont{W.}~\bibnamefont{D\"ur}},
  \bibinfo{author}{\bibfnamefont{G.}~\bibnamefont{Vidal}}, \bibnamefont{and}
  \bibinfo{author}{\bibfnamefont{J.~I.} \bibnamefont{Cirac}},
  \bibinfo{journal}{Physical Review A} \textbf{\bibinfo{volume}{62}},
  \bibinfo{pages}{062314} (\bibinfo{year}{2000}).
\bibinfo{author}{\bibfnamefont{F.}~\bibnamefont{Verstraete}},
  \bibinfo{author}{\bibfnamefont{J.}~\bibnamefont{Dehaene}},
  \bibinfo{author}{\bibfnamefont{B.}~\bibnamefont{De~Moor}}, \bibnamefont{and}
  \bibinfo{author}{\bibfnamefont{H.}~\bibnamefont{Verschelde}},
  \bibinfo{journal}{Physical Review A} \textbf{\bibinfo{volume}{65}},
  \bibinfo{pages}{052112} (\bibinfo{year}{2002}).


\bibitem[{\citenamefont{Bennett et~al.}(1993)\citenamefont{Bennett, Brassard,
  Cr{\'e}peau, Jozsa, Peres, and Wootters}}]{BBC93}
\bibinfo{author}{\bibfnamefont{C.H.}~\bibnamefont{Bennett}},
  \bibinfo{author}{\bibfnamefont{G.}~\bibnamefont{Brassard}},
  \bibinfo{author}{\bibfnamefont{C.}~\bibnamefont{Cr{\'e}peau}},
  \bibinfo{author}{\bibfnamefont{R.}~\bibnamefont{Jozsa}},
  \bibinfo{author}{\bibfnamefont{A.}~\bibnamefont{Peres}}, \bibnamefont{and}
  \bibinfo{author}{\bibfnamefont{W.K.}~\bibnamefont{Wootters}},
  \bibinfo{journal}{Physical Review Letters} \textbf{\bibinfo{volume}{70}},
  \bibinfo{pages}{1895} (\bibinfo{year}{1993}).


\bibitem[{\citenamefont{Briegel et~al.}(1998)\citenamefont{Briegel, D\"ur,
  Cirac, and Zoller}}]{BDC98}
\bibinfo{author}{\bibfnamefont{H.-J.} \bibnamefont{Briegel}},
  \bibinfo{author}{\bibfnamefont{W.}~\bibnamefont{D\"ur}},
  \bibinfo{author}{\bibfnamefont{J.~I.} \bibnamefont{Cirac}}, \bibnamefont{and}
  \bibinfo{author}{\bibfnamefont{P.}~\bibnamefont{Zoller}},
  \bibinfo{journal}{Physical Review Letters} \textbf{\bibinfo{volume}{81}},
  \bibinfo{pages}{5932} (\bibinfo{year}{1998}).

\bibitem[{\citenamefont{Jacobs et~al.}(2002)\citenamefont{Jacobs, Pittman, and
  Franson}}]{JPF02}
\bibinfo{author}{\bibfnamefont{B.~C.} \bibnamefont{Jacobs}},
  \bibinfo{author}{\bibfnamefont{T.~B.} \bibnamefont{Pittman}},
  \bibnamefont{and} \bibinfo{author}{\bibfnamefont{J.~D.}
  \bibnamefont{Franson}}, \bibinfo{journal}{Physical Review A}
  \textbf{\bibinfo{volume}{66}}, \bibinfo{pages}{052307}
  (\bibinfo{year}{2002}). 
\bibinfo{author}{\bibfnamefont{D.}~\bibnamefont{Collins}},
  \bibinfo{author}{\bibfnamefont{N.}~\bibnamefont{Gisin}}, \bibnamefont{and}
  \bibinfo{author}{\bibfnamefont{H.}~\bibnamefont{De~Riedmatten}},
  \bibinfo{journal}{Journal of Modern Optics} \textbf{\bibinfo{volume}{52}},
  \bibinfo{pages}{735} (\bibinfo{year}{2005}).


\bibitem[{\citenamefont{Gingrich et~al.}(2003)\citenamefont{Gingrich, Kok, Lee,
  Vatan, and Dowling}}]{GKL03}
\bibinfo{author}{\bibfnamefont{R.~M.} \bibnamefont{Gingrich}},
  \bibinfo{author}{\bibfnamefont{P.}~\bibnamefont{Kok}},
  \bibinfo{author}{\bibfnamefont{H.}~\bibnamefont{Lee}},
  \bibinfo{author}{\bibfnamefont{F.}~\bibnamefont{Vatan}}, \bibnamefont{and}
  \bibinfo{author}{\bibfnamefont{J.~P.} \bibnamefont{Dowling}},
  \bibinfo{journal}{Physical Review Letters} \textbf{\bibinfo{volume}{91}},
  \bibinfo{pages}{217901} (\bibinfo{year}{2003}).

\bibitem[{\citenamefont{de~Riedmatten et~al.}(2004)\citenamefont{de~Riedmatten,
  Marcikic, Tittel, Zbinden, Collins, and Gisin}}]{DMT04}
\bibinfo{author}{\bibfnamefont{H.}~\bibnamefont{de~Riedmatten}},
  \bibinfo{author}{\bibfnamefont{I.}~\bibnamefont{Marcikic}},
  \bibinfo{author}{\bibfnamefont{W.}~\bibnamefont{Tittel}},
  \bibinfo{author}{\bibfnamefont{H.}~\bibnamefont{Zbinden}},
  \bibinfo{author}{\bibfnamefont{D.}~\bibnamefont{Collins}}, \bibnamefont{and}
  \bibinfo{author}{\bibfnamefont{N.}~\bibnamefont{Gisin}},
  \bibinfo{journal}{Physical Review Letters} \textbf{\bibinfo{volume}{92}},
  \bibinfo{eid}{047904} (\bibinfo{year}{2004}).

\bibitem[{\citenamefont{Cirac and Zoller}(2004)}]{CZ04}
\bibinfo{author}{\bibfnamefont{J.}~\bibnamefont{Cirac}} \bibnamefont{and}
  \bibinfo{author}{\bibfnamefont{P.}~\bibnamefont{Zoller}},
  \bibinfo{journal}{Physics Today} \textbf{\bibinfo{volume}{57}},
  \bibinfo{pages}{38} (\bibinfo{year}{2004}).

\bibitem[{\citenamefont{Jang}(1999)}]{Jan99}
\bibinfo{author}{\bibfnamefont{J.-S.} \bibnamefont{Jang}},
  \bibinfo{journal}{{Physical Review A}} \textbf{\bibinfo{volume}{59}},
  \bibinfo{pages}{2322} (\bibinfo{year}{1999}).
\bibinfo{author}{\bibfnamefont{G.}~\bibnamefont{Krenn}},
  \bibinfo{author}{\bibfnamefont{J.}~\bibnamefont{Summhammer}},
  \bibnamefont{and} \bibinfo{author}{\bibfnamefont{K.}~\bibnamefont{Svozil}},
  \bibinfo{journal}{{Physical Review A}} \textbf{\bibinfo{volume}{61}},
  \bibinfo{pages}{052102} (\bibinfo{year}{2000}).
\bibinfo{author}{\bibfnamefont{G.}~\bibnamefont{Mitchison}} \bibnamefont{and}
  \bibinfo{author}{\bibfnamefont{S.}~\bibnamefont{Massar}},
  \bibinfo{journal}{{Physical Review A}} \textbf{\bibinfo{volume}{63}},
  \bibinfo{pages}{032105} (\bibinfo{year}{2001}).
\bibinfo{author}{\bibfnamefont{H.}~\bibnamefont{Azuma}},
  \bibinfo{journal}{Physical Review A} \textbf{\bibinfo{volume}{74}},
  \bibinfo{eid}{054301} (\bibinfo{year}{2006}).

\bibitem[{\citenamefont{Lanco et~al.}(2006)\citenamefont{Lanco, Ducci,
  Likforman, Marcadet, van Houwelingen, Zbinden, Leo, and Berger}}]{LDL06}
\bibinfo{author}{\bibfnamefont{L.}~\bibnamefont{Lanco}},
  \bibinfo{author}{\bibfnamefont{S.}~\bibnamefont{Ducci}},
  \bibinfo{author}{\bibfnamefont{J.-P.} \bibnamefont{Likforman}},
  \bibinfo{author}{\bibfnamefont{X.}~\bibnamefont{Marcadet}},
  \bibinfo{author}{\bibfnamefont{J.~A.~W.} \bibnamefont{van Houwelingen}},
  \bibinfo{author}{\bibfnamefont{H.}~\bibnamefont{Zbinden}},
  \bibinfo{author}{\bibfnamefont{G.}~\bibnamefont{Leo}}, \bibnamefont{and}
  \bibinfo{author}{\bibfnamefont{V.}~\bibnamefont{Berger}},
  \bibinfo{journal}{Physical Review Letters} \textbf{\bibinfo{volume}{97}},
  \bibinfo{eid}{173901} (\bibinfo{year}{2006}).
\bibinfo{author}{\bibfnamefont{A.}~\bibnamefont{{Badolato}}},
  \bibinfo{author}{\bibfnamefont{K.}~\bibnamefont{{Hennessy}}},
  \bibinfo{author}{\bibfnamefont{M.}~\bibnamefont{{Atat{\"u}re}}},
  \bibinfo{author}{\bibfnamefont{J.}~\bibnamefont{{Dreiser}}},
  \bibinfo{author}{\bibfnamefont{E.}~\bibnamefont{{Hu}}},
  \bibinfo{author}{\bibfnamefont{P.~M.} \bibnamefont{{Petroff}}},
  \bibnamefont{and} \bibinfo{author}{\bibfnamefont{A.}~\bibnamefont{{Imamo{\u
  g}lu}}}, \bibinfo{journal}{Science} \textbf{\bibinfo{volume}{308}},
  \bibinfo{pages}{1158} (\bibinfo{year}{2005}).
\bibinfo{author}{\bibfnamefont{S.}~\bibnamefont{Seidelin}},
  \bibinfo{author}{\bibfnamefont{J.}~\bibnamefont{Chiaverini}},
  \bibinfo{author}{\bibfnamefont{R.}~\bibnamefont{Reichle}},
  \bibinfo{author}{\bibfnamefont{J.~J.} \bibnamefont{Bollinger}},
  \bibinfo{author}{\bibfnamefont{D.}~\bibnamefont{Leibfried}},
  \bibinfo{author}{\bibfnamefont{J.}~\bibnamefont{Britton}},
  \bibinfo{author}{\bibfnamefont{J.~H.} \bibnamefont{Wesenberg}},
  \bibinfo{author}{\bibfnamefont{R.~B.} \bibnamefont{Blakestad}},
  \bibinfo{author}{\bibfnamefont{R.~J.} \bibnamefont{Epstein}},
  \bibinfo{author}{\bibfnamefont{D.~B.} \bibnamefont{Hume}},
  \bibnamefont{et~al.}, \bibinfo{journal}{Physical Review Letters}
  \textbf{\bibinfo{volume}{96}}, \bibinfo{eid}{253003} (\bibinfo{year}{2006}).
\bibinfo{author}{\bibfnamefont{P.}~\bibnamefont{Horak}},
  \bibinfo{author}{\bibfnamefont{B.~G.} \bibnamefont{Klappauf}},
  \bibinfo{author}{\bibfnamefont{A.}~\bibnamefont{Haase}},
  \bibinfo{author}{\bibfnamefont{R.}~\bibnamefont{Folman}},
  \bibinfo{author}{\bibfnamefont{J.}~\bibnamefont{Schmiedmayer}},
  \bibinfo{author}{\bibfnamefont{P.}~\bibnamefont{Domokos}}, \bibnamefont{and}
  \bibinfo{author}{\bibfnamefont{E.~A.} \bibnamefont{Hinds}},
  \bibinfo{journal}{Physical Review A} \textbf{\bibinfo{volume}{67}},
  \bibinfo{pages}{043806} (\bibinfo{year}{2003}).

\bibitem[{\citenamefont{Rudolph}(2000)}]{Rud00}
\bibinfo{author}{\bibfnamefont{T.}~\bibnamefont{Rudolph}},
  \bibinfo{journal}{Physical Review Letters} \textbf{\bibinfo{volume}{85}},
  \bibinfo{pages}{2925} (\bibinfo{year}{2000}).

\bibitem[{\citenamefont{Sondermann et~al.}(2007)\citenamefont{Sondermann,
  Maiwald, Konermann, Lindlein, Peschel, and Leuchs}}]{SMK07}
\bibinfo{author}{\bibfnamefont{M.}~\bibnamefont{Sondermann}},
  \bibinfo{author}{\bibfnamefont{R.}~\bibnamefont{Maiwald}},
  \bibinfo{author}{\bibfnamefont{H.}~\bibnamefont{Konermann}},
  \bibinfo{author}{\bibfnamefont{N.}~\bibnamefont{Lindlein}},
  \bibinfo{author}{\bibfnamefont{U.}~\bibnamefont{Peschel}}, \bibnamefont{and}
  \bibinfo{author}{\bibfnamefont{G.}~\bibnamefont{Leuchs}},
  \bibinfo{journal}{Applied Physics B: Lasers and Optics}
  \textbf{\bibinfo{volume}{89}}, \bibinfo{pages}{489} (\bibinfo{year}{2007}).
\bibinfo{author}{\bibfnamefont{G.}~\bibnamefont{Zumofen}},
  \bibinfo{author}{\bibfnamefont{N.~M.} \bibnamefont{Mojarad}},
  \bibinfo{author}{\bibfnamefont{V.}~\bibnamefont{Sandoghdar}},
  \bibnamefont{and} \bibinfo{author}{\bibfnamefont{M.}~\bibnamefont{Agio}},
  \bibinfo{journal}{Physical Review Letters} \textbf{\bibinfo{volume}{101}},
  \bibinfo{pages}{180404} (\bibinfo{year}{2008}).
\bibinfo{author}{\bibfnamefont{M.~K.} \bibnamefont{Tey}},
  \bibinfo{author}{\bibfnamefont{Z.}~\bibnamefont{Chen}},
  \bibinfo{author}{\bibfnamefont{S.~A.} \bibnamefont{Aljunid}},
  \bibinfo{author}{\bibfnamefont{B.}~\bibnamefont{Chng}},
  \bibinfo{author}{\bibfnamefont{F.}~\bibnamefont{Huber}},
  \bibinfo{author}{\bibfnamefont{G.}~\bibnamefont{Maslennikov}},
  \bibnamefont{and}
  \bibinfo{author}{\bibfnamefont{C.}~\bibnamefont{Kurtsiefer}},
  \bibinfo{journal}{Nature Physics} \textbf{\bibinfo{volume}{4}},
  \bibinfo{pages}{924} (\bibinfo{year}{2008}).

\bibitem[{\citenamefont{Gorshkov et~al.}(2007{\natexlab{a}})\citenamefont{Gorshkov, Andr\'{e}, Fleischhauer,
  Sorensen, and Lukin}}]{GAF07}
\bibinfo{author}{\bibfnamefont{A.~V.} \bibnamefont{Gorshkov}},
  \bibinfo{author}{\bibfnamefont{A.}~\bibnamefont{Andr\'{e}}},
  \bibinfo{author}{\bibfnamefont{M.}~\bibnamefont{Fleischhauer}},
  \bibinfo{author}{\bibfnamefont{A.~S.} \bibnamefont{S{\o}rensen}},
  \bibnamefont{and} \bibinfo{author}{\bibfnamefont{M.~D.} \bibnamefont{Lukin}},
  \bibinfo{journal}{Physical Review Letters} \textbf{\bibinfo{volume}{98}},
  \bibinfo{pages}{123601} (\bibinfo{year}{2007}{\natexlab{a}}).

\bibitem[{\citenamefont{Novikova et~al.}(2007)\citenamefont{Novikova, Gorshkov,
  Phillips, Sorensen, Lukin, and Walsworth}}]{NGP07}
\bibinfo{author}{\bibfnamefont{I.}~\bibnamefont{Novikova}},
  \bibinfo{author}{\bibfnamefont{A.~V.} \bibnamefont{Gorshkov}},
  \bibinfo{author}{\bibfnamefont{D.~F.} \bibnamefont{Phillips}},
  \bibinfo{author}{\bibfnamefont{A.~S.} \bibnamefont{S{\o}rensen}},
  \bibinfo{author}{\bibfnamefont{M.~D.} \bibnamefont{Lukin}}, \bibnamefont{and}
  \bibinfo{author}{\bibfnamefont{R.~L.} \bibnamefont{Walsworth}},
  \bibinfo{journal}{Physical Review Letters} \textbf{\bibinfo{volume}{98}},
  \bibinfo{eid}{243602} (\bibinfo{year}{2007}).


\bibitem[{\citenamefont{Bollinger et~al.}(1996)\citenamefont{Bollinger, Itano,
  Wineland, and Heinzen}}]{BIW96}
\bibinfo{author}{\bibfnamefont{J.~J.} \bibnamefont{Bollinger}},
  \bibinfo{author}{\bibfnamefont{W.~M.} \bibnamefont{Itano}},
  \bibinfo{author}{\bibfnamefont{D.~J.} \bibnamefont{Wineland}},
  \bibnamefont{and} \bibinfo{author}{\bibfnamefont{D.~J.}
  \bibnamefont{Heinzen}}, \bibinfo{journal}{Physical Review A}
  \textbf{\bibinfo{volume}{54}}, \bibinfo{pages}{R4649} (\bibinfo{year}{1996}).
\bibinfo{author}{\bibfnamefont{D.}~\bibnamefont{Leibfried}},
  \bibinfo{author}{\bibfnamefont{M.}~\bibnamefont{Barrett}},
  \bibinfo{author}{\bibfnamefont{T.}~\bibnamefont{Schaetz}},
  \bibinfo{author}{\bibfnamefont{J.}~\bibnamefont{Britton}},
  \bibinfo{author}{\bibfnamefont{J.}~\bibnamefont{Chiaverini}},
  \bibinfo{author}{\bibfnamefont{W.}~\bibnamefont{Itano}},
  \bibinfo{author}{\bibfnamefont{J.}~\bibnamefont{Jost}},
  \bibinfo{author}{\bibfnamefont{C.}~\bibnamefont{Langer}}, \bibnamefont{and}
  \bibinfo{author}{\bibfnamefont{D.}~\bibnamefont{Wineland}},
  \bibinfo{journal}{Science} \textbf{\bibinfo{volume}{304}},
  \bibinfo{pages}{1476} (\bibinfo{year}{2004}).


\bibitem[{\citenamefont{Lukin et~al.}(2001)\citenamefont{Lukin, Fleischhauer,
  Cote, Duan, Jaksch, Cirac, and Zoller}}]{LFC01}
\bibinfo{author}{\bibfnamefont{M.~D.} \bibnamefont{Lukin}},
  \bibinfo{author}{\bibfnamefont{M.}~\bibnamefont{Fleischhauer}},
  \bibinfo{author}{\bibfnamefont{R.}~\bibnamefont{Cote}},
  \bibinfo{author}{\bibfnamefont{L.~M.} \bibnamefont{Duan}},
  \bibinfo{author}{\bibfnamefont{D.}~\bibnamefont{Jaksch}},
  \bibinfo{author}{\bibfnamefont{J.~I.} \bibnamefont{Cirac}}, \bibnamefont{and}
  \bibinfo{author}{\bibfnamefont{P.}~\bibnamefont{Zoller}},
  \bibinfo{journal}{Physical Review Letters} \textbf{\bibinfo{volume}{87}},
  \bibinfo{pages}{037901} (\bibinfo{year}{2001}).

\bibitem[{\citenamefont{Muller et~al.}(2008)\citenamefont{Muller, Lesanovsky,
  Weimer, Buchler, and Zoller}}]{MLW08}
\bibinfo{author}{\bibfnamefont{M.}~\bibnamefont{Muller}},
  \bibinfo{author}{\bibfnamefont{I.}~\bibnamefont{Lesanovsky}},
  \bibinfo{author}{\bibfnamefont{H.}~\bibnamefont{Weimer}},
  \bibinfo{author}{\bibfnamefont{H.~P.} \bibnamefont{Buchler}},
  \bibnamefont{and} \bibinfo{author}{\bibfnamefont{P.}~\bibnamefont{Zoller}},
 \bibinfo{journal}{arXiv:0811.1155v1} (\bibinfo{year}{2008}).

\bibitem[{\citenamefont{Saffman and Walker}(2005)}]{SW08}
\bibinfo{author}{\bibfnamefont{M.}~\bibnamefont{Saffman}} \bibnamefont{and}
  \bibinfo{author}{\bibfnamefont{T.~G.} \bibnamefont{Walker}},
  \bibinfo{journal}{Physical Review A} \textbf{\bibinfo{volume}{72}},
  \bibinfo{pages}{042302} (\bibinfo{year}{2005}).

\bibitem[{\citenamefont{Vogt et~al.}(2006)\citenamefont{Vogt, Viteau, Zhao,
  Chotia, Comparat, and Pillet}}]{VVZ06}
\bibinfo{author}{\bibfnamefont{T.}~\bibnamefont{Vogt}},
  \bibinfo{author}{\bibfnamefont{M.}~\bibnamefont{Viteau}},
  \bibinfo{author}{\bibfnamefont{J.}~\bibnamefont{Zhao}},
  \bibinfo{author}{\bibfnamefont{A.}~\bibnamefont{Chotia}},
  \bibinfo{author}{\bibfnamefont{D.}~\bibnamefont{Comparat}}, \bibnamefont{and}
  \bibinfo{author}{\bibfnamefont{P.}~\bibnamefont{Pillet}},
  \bibinfo{journal}{Physical Review Letters} \textbf{\bibinfo{volume}{97}},
  \bibinfo{pages}{083003} (\bibinfo{year}{2006}).
\bibinfo{author}{\bibfnamefont{R.}~\bibnamefont{Heidemann}},
  \bibinfo{author}{\bibfnamefont{U.}~\bibnamefont{Raitzsch}},
  \bibinfo{author}{\bibfnamefont{V.}~\bibnamefont{Bendkowsky}},
  \bibinfo{author}{\bibfnamefont{B.}~\bibnamefont{Butscher}},
  \bibinfo{author}{\bibfnamefont{R.}~\bibnamefont{L{\"o}w}},
  \bibinfo{author}{\bibfnamefont{L.}~\bibnamefont{Santos}}, \bibnamefont{and}
  \bibinfo{author}{\bibfnamefont{T.}~\bibnamefont{Pfau}},
  \bibinfo{journal}{Physical Review Letters} \textbf{\bibinfo{volume}{99}},
  \bibinfo{pages}{163601} (\bibinfo{year}{2007}).
\bibinfo{author}{\bibfnamefont{M.}~\bibnamefont{Reetz-Lamour}},
  \bibinfo{author}{\bibfnamefont{T.}~\bibnamefont{Amthor}},
  \bibinfo{author}{\bibfnamefont{J.}~\bibnamefont{Deiglmayr}},
  \bibnamefont{and}
  \bibinfo{author}{\bibfnamefont{M.}~\bibnamefont{Weidem{\"u}ller}},
  \bibinfo{journal}{Physical Review Letters} \textbf{\bibinfo{volume}{100}},
  \bibinfo{pages}{253001} (\bibinfo{year}{2008}).
\bibinfo{author}{\bibfnamefont{T.~A.} \bibnamefont{{Johnson}}},
  \bibinfo{author}{\bibfnamefont{E.}~\bibnamefont{{Urban}}},
  \bibinfo{author}{\bibfnamefont{T.}~\bibnamefont{{Henage}}},
  \bibinfo{author}{\bibfnamefont{L.}~\bibnamefont{{Isenhower}}},
  \bibinfo{author}{\bibfnamefont{D.~D.} \bibnamefont{{Yavuz}}},
  \bibinfo{author}{\bibfnamefont{T.~G.} \bibnamefont{{Walker}}},
  \bibnamefont{and}
  \bibinfo{author}{\bibfnamefont{M.}~\bibnamefont{{Saffman}}},
  \bibinfo{journal}{Physical Review Letters} \textbf{\bibinfo{volume}{100}},
  \bibinfo{pages}{113003} (\bibinfo{year}{2008}).
\bibinfo{author}{\bibfnamefont{E.}~\bibnamefont{Urban}},
  \bibinfo{author}{\bibfnamefont{T.~A.} \bibnamefont{Johnson}},
  \bibinfo{author}{\bibfnamefont{T.}~\bibnamefont{Henage}},
  \bibinfo{author}{\bibfnamefont{L.}~\bibnamefont{Isenhower}},
  \bibinfo{author}{\bibfnamefont{D.~D.} \bibnamefont{Yavuz}},
  \bibinfo{author}{\bibfnamefont{T.~G.} \bibnamefont{Walker}},
  \bibnamefont{and} \bibinfo{author}{\bibfnamefont{M.}~\bibnamefont{Saffman}}
  \bibinfo{journal}{arXiv.org:0805.0758v1} (\bibinfo{year}{2008}).
\bibinfo{author}{\bibfnamefont{A.}~\bibnamefont{Gaëtan}},
  \bibinfo{author}{\bibfnamefont{Y.}~\bibnamefont{Miroshnychenko}},
  \bibinfo{author}{\bibfnamefont{T.}~\bibnamefont{Wilk}},
  \bibinfo{author}{\bibfnamefont{A.}~\bibnamefont{Chotia}},
  \bibinfo{author}{\bibfnamefont{M.}~\bibnamefont{Viteau}},
  \bibinfo{author}{\bibfnamefont{D.}~\bibnamefont{Comparat}},
  \bibinfo{author}{\bibfnamefont{P.}~\bibnamefont{Pillet}},
  \bibinfo{author}{\bibfnamefont{A.}~\bibnamefont{Browaeys}}, \bibnamefont{and}
  \bibinfo{author}{\bibfnamefont{P.}~\bibnamefont{Grangier}}
  \bibinfo{journal}{arXiv.org:0810.2960v1} (\bibinfo{year}{2008}). 


\bibitem[{\citenamefont{Cere et~al.}(2008)\citenamefont{Cere, Parigi, Abad,
  Wolfgramm, Predojevic, and Mitchell}}]{CPA08}
\bibinfo{author}{\bibfnamefont{A.}~\bibnamefont{Cer\`{e}}},
  \bibinfo{author}{\bibfnamefont{V.}~\bibnamefont{Parigi}},
  \bibinfo{author}{\bibfnamefont{M.}~\bibnamefont{Abad}},
  \bibinfo{author}{\bibfnamefont{F.}~\bibnamefont{Wolfgramm}},
  \bibinfo{author}{\bibfnamefont{A.}~\bibnamefont{Predojevic}},
  \bibnamefont{and} \bibinfo{author}{\bibfnamefont{M.}~\bibnamefont{Mitchell}},
  \bibinfo{journal}{arXiv:0812.2326v1}  (\bibinfo{year}{2008}).


\bibitem[{\citenamefont{Franson et~al.}(2002)\citenamefont{Franson, Donegan,
  Fitch, Jacobs, and Pittman}}]{FDF02}
\bibinfo{author}{\bibfnamefont{J.~D.} \bibnamefont{Franson}},
  \bibinfo{author}{\bibfnamefont{M.~M.} \bibnamefont{Donegan}},
  \bibinfo{author}{\bibfnamefont{M.~J.} \bibnamefont{Fitch}},
  \bibinfo{author}{\bibfnamefont{B.~C.} \bibnamefont{Jacobs}},
  \bibnamefont{and} \bibinfo{author}{\bibfnamefont{T.~B.}
  \bibnamefont{Pittman}}, \bibinfo{journal}{{Physical Review Letters}}
  \textbf{\bibinfo{volume}{89}}, \bibinfo{pages}{137901}
  (\bibinfo{year}{2002}).

\bibitem[{\citenamefont{Aliferis and Leung}(2004)}]{AL04}
\bibinfo{author}{\bibfnamefont{P.}~\bibnamefont{Aliferis}} \bibnamefont{and}
  \bibinfo{author}{\bibfnamefont{D.~W.} \bibnamefont{Leung}},
  \bibinfo{journal}{Physical Review A} \textbf{\bibinfo{volume}{70}},
  \bibinfo{eid}{062314} (\bibinfo{year}{2004}).

\bibitem[{\citenamefont{Popescu}(2007{\natexlab{b}})}]{Pop07}
\bibinfo{author}{\bibfnamefont{S.}~\bibnamefont{Popescu}},
  \bibinfo{journal}{Physical Review Letters} \textbf{\bibinfo{volume}{99}},
  \bibinfo{eid}{250501} (\bibinfo{year}{2007}{\natexlab{b}}).


\bibitem[{\citenamefont{{Dicke}}(1964)}]{Dic64}
\bibinfo{author}{\bibfnamefont{R.~H.} \bibnamefont{{Dicke}}}, in
  \emph{\bibinfo{booktitle}{Quantum Electronics}}, edited by
  \bibinfo{editor}{\bibfnamefont{P.}~\bibnamefont{{Grivet}}} \bibnamefont{and}
  \bibinfo{editor}{\bibfnamefont{N.}~\bibnamefont{{Bloembergen}}}
, pp. \bibinfo{pages}{35--53}  (\bibinfo{year}{1964}).


\bibitem[{\citenamefont{Dicke}(1954)}]{Dic54}
\bibinfo{author}{\bibfnamefont{R.~H.} \bibnamefont{Dicke}},
  \bibinfo{journal}{Phys. Rev.} \textbf{\bibinfo{volume}{93}},
  \bibinfo{pages}{99} (\bibinfo{year}{1954}).

\end{thebibliography}
\end{document}